\documentclass[numberedappendix,12pt,preprint]{emulateapj}  
\usepackage[]{amsmath}
\usepackage{graphicx}
\usepackage{multirow}
\usepackage{booktabs}
\usepackage{bm}
\usepackage{xspace}

\usepackage{afterpage}
\usepackage{url}
\usepackage{textcomp}

\usepackage{color}
\usepackage[usenames,dvipsnames]{xcolor}
\definecolor{midgray}{gray}{0.4}
\usepackage[colorlinks=true,citecolor=blue,linkcolor=magenta]{hyperref}

\newcommand{\simgt}{\,\rlap{\lower 3.5 pt \hbox{$\mathchar \sim$}} \raise
1pt \hbox {$>$}\,}
\newcommand{\simlt}{\,\rlap{\lower 3.5 pt \hbox{$\mathchar \sim$}} \raise
1pt \hbox {$<$}\,}

\newcommand{\HST}{\textit{HST}\xspace}

\newcommand{\BE}{\begin{equation}}
\newcommand{\EE}{\end{equation}}
\newcommand{\BEA}{\begin{eqnarray}}
\newcommand{\EEA}{\end{eqnarray}}

\newcommand{\msolar}{M_{\odot}\xspace}

\newcommand{\ssfrunit}{\xspace\mathrm{Gyr}^{-1}\xspace}
\newcommand{\mstar}{M_\star\xspace}
\newcommand{\lya}{Ly$\alpha$\xspace}
\newcommand{\Ha}{H$\alpha$\xspace}
\newcommand{\Hb}{H$\beta$\xspace}

\newcommand{\OII}{[O \textsc{II}]\xspace}

\newcommand{\OIII}{[O \textsc{III}]\xspace}

\newcommand{\degree}{$^{\circ}$\xspace}

\newcommand{\galpak}{GalPaK$^{3D}$\xspace}
\newcommand{\kmosd}{KMOS$^{3D}$\xspace}

\shorttitle{KLASS: Kinematics of Lensed Galaxies at Cosmic Noon}
\shortauthors{Mason et al. (2016)}

\begin{document}
\title{First Results from the KMOS Lens-Amplified Spectroscopic Survey (KLASS): \\ Kinematics of Lensed Galaxies at Cosmic Noon}

\author{
Charlotte~A.~Mason$^{1}$,
Tommaso~Treu$^{1}$,
Adriano~Fontana$^{2}$,
Tucker~Jones$^{3,4,5}$,
Takahiro~Morishita$^{1,6,7}$,
Ricardo~Amorin$^{8,9}$,
Maru\v{s}a~Brada\v{c}$^{4}$,
Emily~Quinn~Finney$^{4}$,
Claudio~Grillo$^{10,11}$,
Alaina~Henry$^{12}$,
Austin~Hoag$^{4}$,
Kuang-Han~Huang$^{4}$,
Kasper~B.~Schmidt$^{13}$,
Michele~Trenti$^{14}$, and
Benedetta~Vulcani$^{14}$
}
\affil{$^{1}$ Department of Physics and Astronomy, University of California, Los Angeles, CA, USA 90095-1547}
\affil{$^{2}$ INAF Osservatorio Astronomico di Roma, Via Frascati 33, 00040 Monteporzio (RM), Italy}
\affil{$^{3}$ Institute of Astronomy, University of Hawaii, 2680 Woodlawn Drive, Honolulu, HI 96822, USA}
\affil{$^{4}$ Department of Physics, University of California, Davis, CA, 95616, USA}
\affil{$^{5}$ Hubble Fellow}
\affil{$^{6}$ Astronomical Institute, Tohoku University, Aramaki, Aoba, Sendai 980-8578, Japan}
\affil{$^{7}$ Institute for International Advanced Research and Education, Tohoku University, Aramaki, Aoba, Sendai 980-8578, Japan}
\affil{$^{8}$ Cavendish Laboratory, University of Cambridge, 19 J.J. Thomson Ave., Cambridge CB30HE, UK}
\affil{$^{9}$ Kavli Institute for Cosmology, University of Cambridge, Madingley Road, Cambridge, CB30HA, UK}
\affil{$^{10}$ Dipartimento di Fisica, Universit\`a  degli Studi di Milano, via Celoria 16, I-20133 Milano, Italy}
\affil{$^{11}$ Dark Cosmology Centre, Niels Bohr Institute, University of Copenhagen, Juliane Maries Vej 30, DK-2100 Copenhagen, Denmark}
\affil{$^{12}$ Space Telescope Science Institute, 3700 San Martin Drive, Baltimore, MD, 21218, USA}
\affil{$^{13}$ Leibniz-Institut f\"uŸr Astrophysik Potsdam (AIP), An der Sternwarte 16, 14482 Potsdam, Germany}
\affil{$^{14}$ School of Physics, University of Melbourne, VIC 3010, Australia}
\email{cmason@astro.ucla.edu}

\begin{abstract}
We present the first results of the KMOS Lens-Amplified Spectroscopic Survey (KLASS), a new ESO Very Large Telescope (VLT) large program, doing multi-object integral field spectroscopy of galaxies gravitationally lensed behind seven galaxy clusters selected from the \HST Grism Lens-Amplified Survey from Space (GLASS). Using the power of the cluster magnification we are able to reveal the kinematic structure of 25 galaxies at $0.7 \simlt z \simlt 2.3$, in four cluster fields, with stellar masses $7.8 \simlt \log{(M_\star/M_\odot)} \simlt 10.5$. This sample includes 5 sources at $z>1$ with lower stellar masses than in any previous kinematic IFU surveys. Our sample displays a diversity in kinematic structure over this mass and redshift range. The majority of our kinematically resolved sample is rotationally supported, but with a lower ratio of rotational velocity to velocity dispersion than in the local universe, indicating the fraction of dynamically hot disks changes with cosmic time. We find no galaxies with stellar mass $<3 \times 10^9 M_\odot$ in our sample display regular ordered rotation. Using the enhanced spatial resolution from lensing, we resolve a lower number of dispersion dominated systems compared to field surveys, competitive with findings from surveys using adaptive optics. We find that the KMOS IFUs recover emission line flux from \HST grism-selected objects more faithfully than slit spectrographs. With artificial slits we estimate slit spectrographs miss on average 60\% of the total flux of emission lines, which decreases rapidly if the emission line is spatially offset from the continuum.
\end{abstract}

\section{Introduction}

With the advent of integral field spectroscopy, which obtain spectra in spatial pixels, it is finally possible to achieve a three-dimensional view of galaxies. Spatially resolved spectroscopy allows us to observe large star-forming regions themselves and make inferences about the physical conditions within galaxies. 

The redshift range $1\simlt z\simlt3$ was the most active time in the universe's history, covering the peak of the cosmic star formation history~\citep{Madau2014} when more than half of the stellar mass in the universe was built up \citep{Ilbert2013,Muzzin2013}. Photometric surveys have revealed that star formation rates (SFRs) and SFR surface densities in this period are systematically higher than in the local universe \citep[e.g.,][]{Madau1996,Hopkins2006,Willott2015,Shibuya2015}. Many galaxies at this epoch appear morphologically disordered \citep[e.g.][]{Mortlock2013,Lee2013,Shapley2001}, a far cry from the clear morphological bimodality in the galaxy population in the local universe, between rotating disks and dispersion dominated elliptical galaxies. How this bimodality arises and what processes change galaxies from disks to ellipticals are still open questions \citep{Conselice2014,Bundy2005}. Merger interactions are expected to play a role in shaping galaxies \citep{Nipoti2003,Bundy2005,Puech2012}, but observing such dynamical processes via the `snapshots' available to astronomers is challenging.

Using integral field spectroscopy we can ask questions about how galaxies' morphologies and kinematics are related to their past and ongoing star formation. A key question is whether the increase in SFRs is purely driven by an increase in density and smooth gas accretion rates at higher redshifts \citep[e.g.,][]{Mason2015a,Tacchella2013} producing steady in-situ star formation \citep{Bundy2007,Conselice2014}, or more stochastic processes for gas infall such as major mergers \citep{Somerville2001,Cole2002}. Additionally, changing physical conditions at high redshift may alter the nature and efficiency of star formation: e.g. decreased AGN activity, lower metallicities, or other evolving feedback processes \citep{Hayward2015,Cullen2016}.

The first generations of integral fields surveys using single integral field unit (IFU) instruments, e.g. SINFONI/SINS \citep{ForsterSchreiber2006,Genzel2011}, SINFONI/AMAZE-LSD \citep{Gnerucci2010} and IMAGES/FLAMES-GIRAFFE \citep{Flores2006} have primarily targeted star forming galaxies with stellar masses $\simgt 10^{10} M_\odot$. Most surveys found samples of $z\sim1-3$ galaxies which were roughly equally separated into 3 kinematic classifications: rotation dominated systems, dispersion dominated systems, and merging/morphological unstable systems. A key result was that the rotation dominated systems had systematically higher velocity dispersions than local disks \citep{Epinat2010,Bershady2010}, suggesting that high redshifts disks are highly turbulent. In addition, the highest mass objects were rotating disks at high redshift, in contrast to the local universe where most objects with stellar mass over $10^{10} M_\odot$ are dispersion dominated ellipticals. Similar trends are seen with slit spectrographs \citep[e.g.][]{Price2016}. 

The K-band Multi-Object Spectrometer \citep[KMOS,][]{Sharples2013} on the European Southern Observatories Very Large Telescope (ESO/VLT) is the first multi-object near-IR IFU instrument and capable of producing large samples of kinematically resolved galaxies. Recent surveys using KMOS, \kmosd \citep{Wisnioski2015} and KROSS \citep{Stott2016}, find the majority of \Ha-selected galaxies at $z\sim1-2$ are highly turbulent gas rich disks. However, these and previous kinematic surveys have probed only the high mass end of the galaxy mass function ($\simgt 10^{9} M_\odot$) and thus there is no clear picture of the kinematic evolution of low mass galaxies. 

Seeing-limited IFU observations have been shown to misclassify objects: at low spatial resolution beam smearing can both smooth out irregular rotation curves so that kinematically irregular galaxies look like rotators \citep{Leethochawalit2016}, or produce large velocity dispersions in kinematic maps of compact galaxies, so that rotators look like dispersion dominated systems \citep{Newman2013a}. Adaptive optics (AO) on the single-object IFU instruments Keck/OSIRIS and VLT/SINFONI have enabled high spatial resolution spectroscopy of a handful of $z\sim1-3$ galaxies \citep{Newman2013a}, including objects which are gravitationally lensed, with stellar masses as low as $6.3 \times 10^8 M_\odot$ \citep{Jones2010a,Livermore2015,Leethochawalit2016}. The surveys using AO find a lower fraction of dispersion dominated systems: high spatial resolution is needed to clearly distinguish rotationally supported galaxies from mergers and pressure supported systems.

Whilst IFU surveys at $z\simgt1$ have produced interesting results, there are clear limitations: there is a need for large samples of galaxies, spanning a broad range in stellar mass, and with higher spatial resolution than provided by natural seeing. Gravitational lensing provides a unique tool to study the internal motions of galaxies with lower stellar masses than in the field, and at higher spatial resolution than natural seeing. By targeting cluster lens fields with a multi-object IFU instrument such as KMOS we can efficiently produce a large sample of lensed high redshift star forming galaxies for the first time.

The KMOS Lens-Amplified Spectroscopic Survey (KLASS) was designed to efficiently survey lensed low mass high redshift galaxies in order to answer questions about how galaxies' star formation histories are related to their kinematics. KLASS is an ESO/VLT large program (PI: A. Fontana), targeting gravitationally lensed galaxies behind seven massive clusters from the Grism Lens Amplified Survey from Space \citep[GLASS, PI: T. Treu,][]{Treu2015,Schmidt2014}. 

In this paper we present a kinematic study of 32 lensed galaxies at cosmic noon ($1\simlt z\simlt 3$) with stellar masses $7.8 \simlt \log{(M_\star/M_\odot)} \simlt 10.5$. We resolve kinematic structure with a high signal-to-noise ratio (S/N $>5$) in 25/32 galaxies. 

By combining the magnifying power of gravitational lensing (median magnification factor of $\sim2$ for the objects presented here) and the multi-object capabilities of KMOS, KLASS efficiently surveys galaxies at better spatial resolution than natural seeing and with stellar mass up to an order of magnitude smaller than previous studies. 5 of the galaxies with resolved kinematics at $z>1$ in our sample have stellar masses below $6.3\times10^8 M_\odot$, lower than any object previously observed with an IFU. 

Our sample reveals a large diversity in the star forming galaxy population at cosmic noon, with a range in inferred kinematic structure and galaxy properties at every redshift.

This paper is structured as follows: in Section~\ref{sec:survey} we introduce the GLASS and KLASS surveys; in Section~\ref{sec:data} we describe the KLASS sample selection, observations and data reduction in Section~\ref{sec:results} we present the analysis and key results of our data, which are discussed in Section~\ref{sec:discussion}. We summarize our findings in Section~\ref{sec:conc}.

We use a \citet{PlanckCollaboration2015} cosmology and all magnitudes are in the AB system.

\section{The KMOS Lens-Amplified Spectroscopic Survey}\label{sec:survey}

KLASS is an ongoing ESO VLT KMOS Large Program targeting the fields of seven massive galaxy clusters, including the four Hubble Frontier Fields visible from the Southern Hemisphere.  A comprehensive description of the survey and data will be presented in Mason et al. (in prep). Here, we provide a brief overview. 

KLASS is a ground-based follow-up program for the Grism Lens Amplified Survey from Space\footnote{\url{http://glass.astro.ucla.edu}} \citep[GLASS,][]{Treu2015,Schmidt2014}, a large Hubble Space Telescope (\HST) program which has obtained grism spectroscopy of the fields of ten massive galaxy clusters, including the Hubble Frontier Fields \citep[HFF,][]{Lotz2016} and 8 of the CLASH clusters \citep{Postman2011}. Near infra-red spectra were obtained with the Wide Field Camera 3 (WFC3) grisms G102 and G141, covering the wavelength range $0.8 - 1.6\mu$m with spectral resolution $R\sim150$. Full details of the GLASS survey are described in \citet{Treu2015} and \citet{Schmidt2014}.

High spectral resolution follow-up is needed confirm the purity and completeness of the grism spectra, to measure lines that were unresolved in \HST, and to obtain velocity information which the low resolution grisms cannot provide.

The key science drivers of KLASS are:
\begin{enumerate}
\item To probe the internal kinematics of galaxies at $z\sim1-3$, with superior spatial resolution to comparable surveys in blank fields. The kinematic data will be combined with metallicity gradients from the \HST data to enable the study of metallicity gradients as a diagnostic of gas inflows and outflows \citep{Jones2013a,Jones2015,Wang2016}
\item To confirm $z\simgt7$ \lya emission from the GLASS sample, enabling us to constrain the timeline and topology of reionization \citep{Schmidt2016,Treu2013,Treu2012}.
\end{enumerate}
The former science driver is the main focus of this paper; the latter will be discussed in a future paper.

In this paper we present the first results for 32 targets, with $\sim10-50\%$ of the planned exposure times, from four of the clusters: MACJS0416.1-2403 (hereafter MACS0416); MACSJ1149.6+2223 (MACS1149); MACSJ2129.4-0741 (MACS2129) and RXJ1347.5-1145 (RXJ1347). When complete, KLASS will have approximately 60 targets at $z\sim1-3$ with all seven clusters. We are targeting $\sim70$ candidate galaxies at $z>7$ which will be described in future work after the full integrations are complete. Final integration times of targets are expected to be $10-15$ hours.

\section{Observations and Data}\label{sec:data}
 
In this section we describe the KMOS observations of the sample presented in this paper.

\subsection{Target selection}\label{sec:data_design}

Targets at `cosmic noon' ($1 \simlt z \simlt 3$) were selected from the \HST GLASS spectroscopic sample, with at least one bright nebular emission line (\Ha, \OIII or \OII) in the KMOS YJ range, away from bright OH sky lines. \HST grism spectra for all of the targets presented in this paper are available in the public GLASS data release \footnote{\url{https://archive.stsci.edu/prepds/glass/}}. The selection by line flux means our sample is comprised of star forming galaxies, which would not necessarily be the case for mass-selected samples.

\subsection{Observations and data reduction}\label{sec:data_reduction}

Observations presented in this paper were carried out in service mode in Periods $95-97$, from July 2015 to April 2016. The KMOS YJ band is used for the entire program, covering $1-1.35\mu$m, with spectral resolution $R\sim3400$, as required to resolve kinematics in our sample. Observations are executed in 1 hour observing blocks (each comprising a total 1800s on science objects and 900s on sky). Pixel dither shifts were included between science frames. A star is observed in 1 IFU in every observing block to monitor the point spread function (PSF) and the accuracy of dither offsets. All exposures had seeing $\leq 0\farcs8$ with median seeing $\sim 0\farcs6$. This corresponds to a spatial resolution of $\sim5/\sqrt{\mu}$ kpc at $z\sim1$, where $\mu$ is the gravitational lensing magnification of an object.

The total integration times of individual objects are listed in Table~\ref{tab:targets}. The integration times are comparable to those of KROSS \citep[2.5 hours per source,][]{Stott2016} and generally lower than \kmosd \citep[2-20 hours per source,][]{Wisnioski2015}. Though we note that the targets should all have 10 hour integrations when KLASS is complete.

All data were reduced using the ESO KMOS pipeline \citep{Davies2013} including an optimized sky subtraction routine from \cite{Davies2007}. We also apply a correction for readout channel level offsets, a known problem with the KMOS detectors. Individual data cube frames are combined by sigma clipping and using spatial shifts determined by the position of the star observed in the same IFU in each frame. A full description of the reduction procedure will be given by Mason et al. (in prep).

\section{Analysis and Results}\label{sec:results}

In this section we present the key analyses and findings of our investigation. We find KMOS measures emission flux consistent with the \HST grisms. After deriving galaxy properties from SED fitting and kinematic modeling, and correcting for lensing effects, we classify our sample in 5 kinematic categories. We find the majority are rotation supported and investigate correlations between kinematic properties and star formation parameters.

\subsection{Comparison of \HST grism and ground-based flux measurements}\label{sec:res_flux}

HST grism surveys such as GLASS and the Faint Infrared Grism Survey \citep[FIGS, PI: Malhotra,][]{Tilvi2016} are providing high spatial resolution near-IR spectroscopy, free from atmospheric attenuation. However, emission lines discovered in the \HST grisms and followed up from the ground have shown some tension in line flux measurements: both for intermediate redshift nebular emission lines \citep{Masters2014} and \lya emission at $z\simgt7$, with Keck/MOSFIRE measuring up to $\sim 5 \times$ lower flux than the \HST G102 grism for \lya \citep[][Hoag et al. in prep]{Tilvi2016}.

Ensuring consistent flux measurements from space and ground-based spectroscopy is important to calibrate instruments, and for making physical inferences from data. Accurate flux measurements of \lya emission are vital for making inferences about the epoch of reionization \citep{Treu2012,Treu2013}, which is a key science driver of KLASS and GLASS. Thus we want to investigate how fluxes measured with KMOS compare with those from the \HST grisms.

\begin{figure}[!t]
\includegraphics[width=0.45\textwidth]{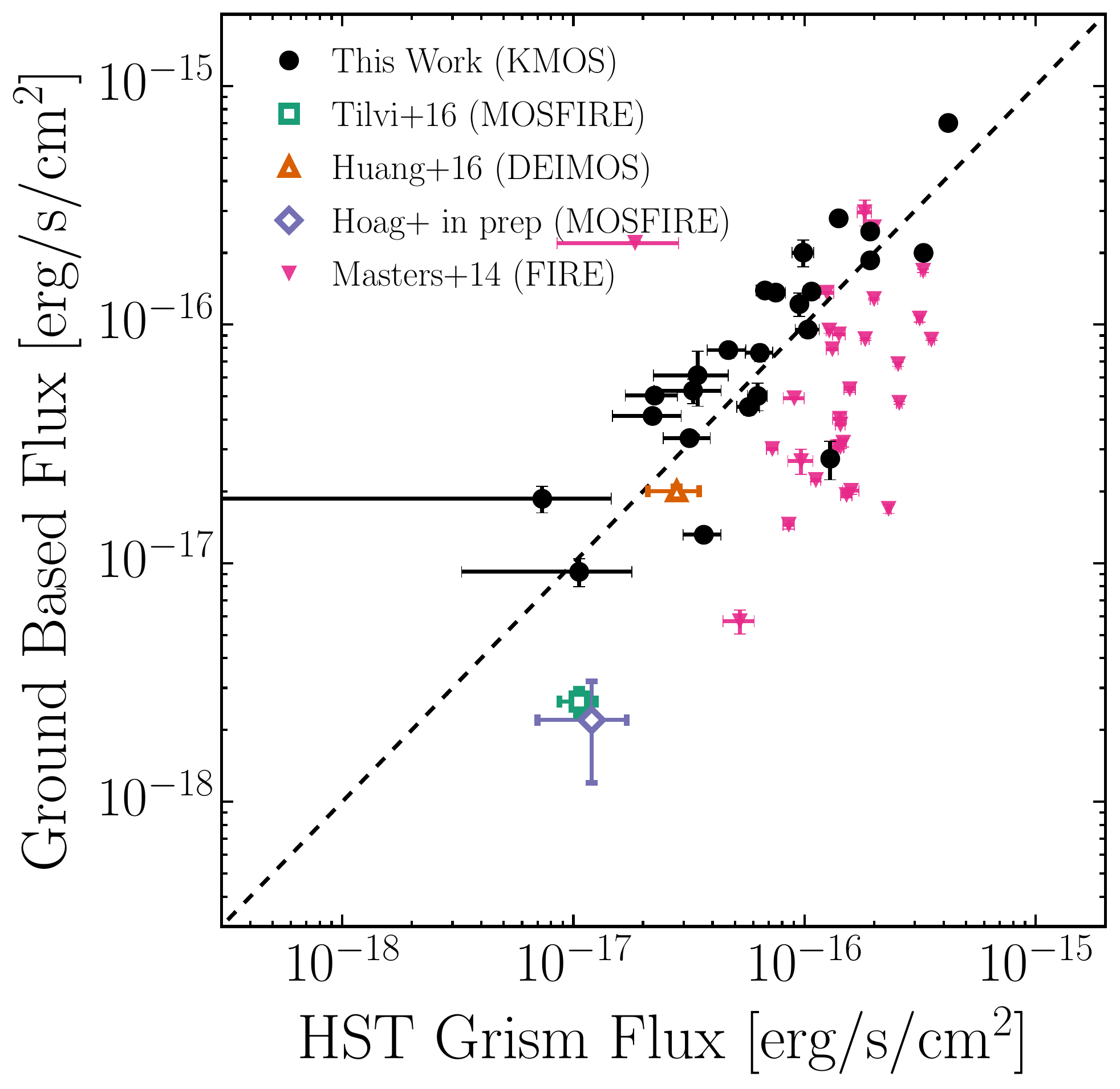}
\caption{Line flux of nebular emission lines from multiple objects in the KLASS first results sample (black points), measured both from the \HST grism in GLASS and from KMOS as described in this paper. We compare the KMOS data with recent results using slit spectrographs, which also have \HST grism spectra \citep[][Hoag et al. in prep]{Masters2014,Tilvi2016,Huang2016a}. Nebular emission line measurements are shown as filled shapes, and \lya line measurements are shown as empty shapes.}
\label{fig:compare_flux}
\end{figure}

In Figure~\ref{fig:compare_flux} we compare flux measured by the \HST grisms G102 and G141 and flux measured by KMOS YJ, for emission lines of objects $1 \simlt z \simlt 3$. We find good agreement between line fluxes measured in KMOS and those from the \HST grisms in GLASS, in contrast to results from slit spectrographs \citep[][Hoag et al. in prep]{Masters2014,Tilvi2016,Huang2016a,Schmidt2016}, suggesting that slit losses can be a serious systematic problem, especially for faint objects, which seems to be avoided by the wide field of IFUs such as KMOS. Keck/DEIMOS and MOSFIRE are traditionally the instrument of choice for following-up faint high redshift targets, but our results show that IFU instruments such as KMOS are necessary to collect the true emission line flux from an object, especially if the emission is likely to be more extended than, and/or offset, from the UV continuum observed with \HST~- as is often the case for highly resonant lines such as \lya \citep{Wisotzki2016,Tilvi2016}, and the offset can be enhanced by lensing \citep{Smit2017}.

To quantify the effect of slit losses and emission lines offset from the continuum we placed artificial slits on the KMOS data cubes. We used a slit width of $0\farcs8 = 4$ pixels, comparable to the slit width of the MOSFIRE/MOSDEF kinematic survey \citep[$0\farcs7$,][]{Price2016}. We placed slits at the peak of the emission line and then at increasing radial offsets from the peak. We measure the flux in the slits aligned at random orientations and take the mean value from all slit orientations at each offset distance.

\begin{figure}[t!],
\includegraphics[width=0.45\textwidth]{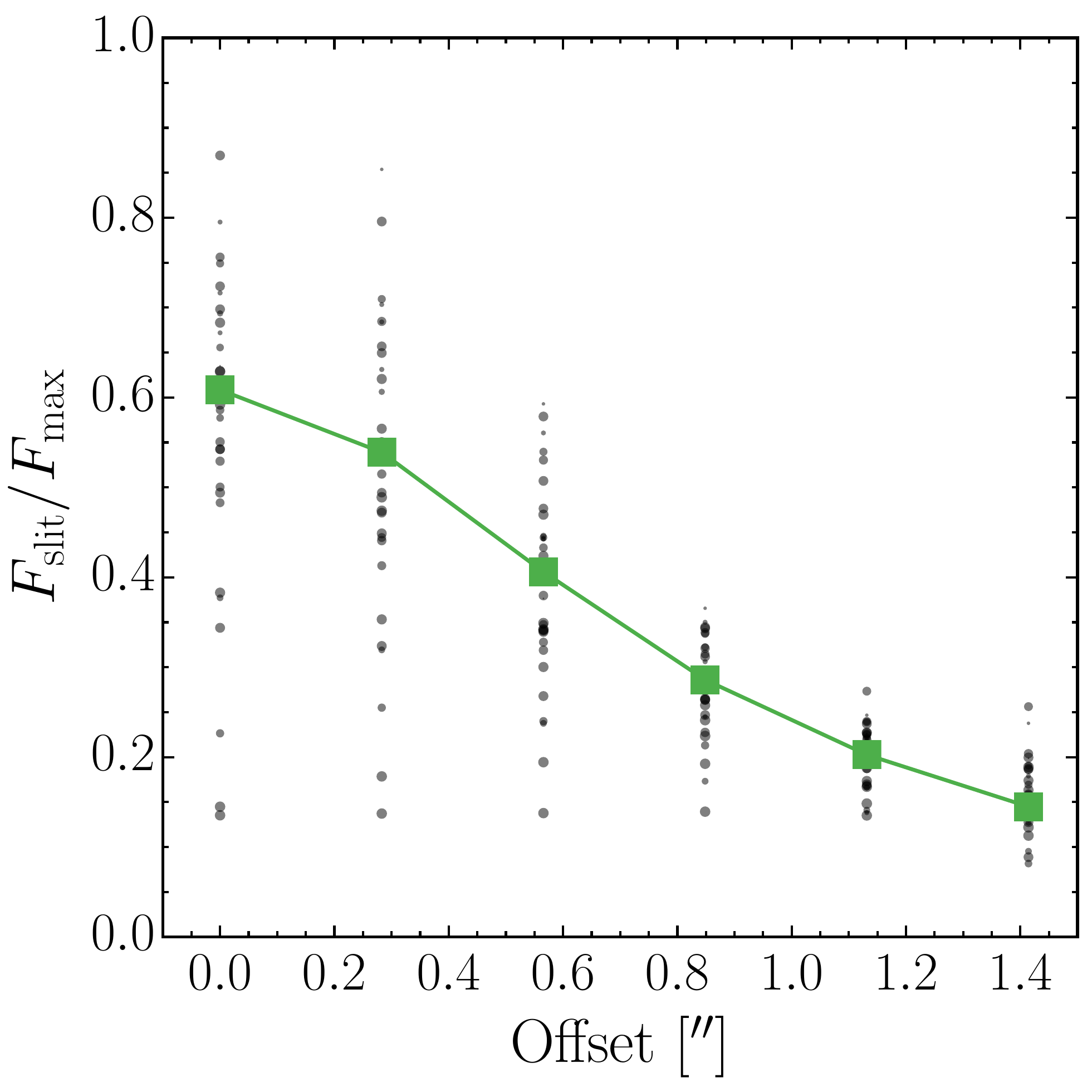}
\caption{Fraction of emission flux measured by artificial slits with width $0\farcs8$ compared to the total flux in a KMOS data cube as a function of slit spatial offset from the emission line center. Gray points are the mean measurement from artificial slits for all emission lines presented in this paper, with size representing the spatial extent of the emission line determined from a S/N map. The green points show the mean of the flux ratio for all lines at each spatial offset value. As well as regular slit losses, if the slit is not centered on an emission line a significant proportion of the total flux will be missed.}
\label{fig:slitloss}
\end{figure}

In Figure~\ref{fig:slitloss} we plot the ratio between flux measured in the slit and the total emission line flux in the KMOS cube as a function of spatial offset. When the spatial offset is $0\arcsec$ we recover the slit loss due to slit size. For our sample, $\sim15-90\%$ of the total flux is measured when using a slit placed on the center of the emission line, with mean value $61\pm1\%$. There is a large scatter in recovered line ratios, due to spatial extent and varying morphologies, this produces ratios comparable to those seen by \citet{Masters2014}. However, if the slit is offset from the spatial center of the emission line, the recovered flux decreases significantly: for a spatial offset of $1\farcs1$ the mean recovered flux ratio is only $\sim20\%$. This flux loss due to spatial offset may explain the tension found for the \lya measurements.

\begin{table*}[t!]
\centering
\caption[ ]{Observed and derived galaxy properties}
\label{tab:targets}
\begin{tabular}[c]{cccccccccccc}
\hline
\hline
Cluster & ID$^a$ & R.A. & Dec. & $z_\textsc{kmos}$ & Kinematic  & Integration   &  Magnification   & $\log_{10}\mstar$ & $\sigma_0$   & $V_\mathrm{max}$ & Kinematic \\
        &        &      &      &                   &  Line      & Time (hrs)      &  $\mu^b$         & $[\msolar]^c$     & [km/s]             & [km/s]     & Class$^d$ \\
\hline 
MACS0416 & 94 & $64.0331$ & $-24.0563$ & $1.37$ & \OIII & $4.75$ & $1.89_{-0.01}^{+0.02}$ & $9.88\pm0.13$ & $<38$ & $31\pm9$ & $3$ \\
MACS0416 & 372 & $64.0352$ & $-24.0710$ & $1.99$ & \OII & $4.75$ & $2.37_{-0.02}^{+0.01}$ & $10.20\pm0.10$ & $43\pm28$ & $-$ & $5$ \\
MACS0416 & 394 & $64.0366$ & $-24.0673$ & $0.94$ & \Ha & $4.75$ & $18.47_{-3.03}^{+0.96}$ & $9.22\pm0.12$ & $12\pm15$ & $133\pm22$ & $2^e$ \\
MACS0416 & 430 & $64.0536$ & $-24.0660$ & $2.10$ & \OII & $4.75$ & $2.92_{-0.05}^{+0.05}$ & $9.17_{-0.18}^{+0.12}$ & $<30$ & $-$ & $5$ \\
MACS0416 & 706 & $64.0514$ & $-24.0713$ & $1.35$ & \OIII & $4.75$ & $2.11_{-0.03}^{+0.03}$ & $8.31\pm0.11$ & $<36$ & $24\pm7$ & $3$ \\
MACS0416 & 863 & $64.0169$ & $-24.0742$ & $1.63$ & \OIII & $4.75$ & $2.98_{-0.11}^{+0.12}$ & $8.84_{-0.81}^{+0.27}$ & $38\pm6$ & $191\pm43$ & $2^e$ \\
MACS0416 & 880 & $64.0310$ & $-24.0790$ & $1.64$ & \OIII & $4.75$ & $2.22_{-0.13}^{+0.14}$ & $9.73\pm0.11$ & $<34$ & $62\pm5$ & $4$ \\
MACS0416 & 955 & $64.0419$ & $-24.0758$ & $1.99$ & \OII & $4.75$ & $3.21_{-0.04}^{+0.07}$ & $9.77\pm0.10$ & $31\pm23$ & $-$ & $5$ \\
MACS1149 & 593 & $177.4069$ & $22.4075$ & $1.48$ & \OIII & $2.25$ & $2.34_{-0.01}^{+0.02}$ & $9.27\pm0.10$ & $8\pm3$ & $142\pm9$ & $1$ \\
MACS1149 & 683 & $177.3972$ & $22.4062$ & $1.68$ & \OIII & $2.25$ & $9.90_{-0.40}^{+0.33}$ & $7.75\pm0.10$ & $<25$ & $30\pm6$ & $4$ \\
MACS1149 & 691 & $177.3824$ & $22.4058$ & $0.98$ & \Ha & $2.25$ & $2.00_{-0.07}^{+0.11}$ & $9.30_{-0.17}^{+0.12}$ & $24\pm11$ & $27\pm14$ & $4$ \\
MACS1149 & 862 & $177.4034$ & $22.4024$ & $1.49$ & \OIII & $2.25$ & $3.90_{-0.04}^{+0.04}$ & $9.66\pm0.10$ & $<34$ & $-$ & $5$ \\
MACS1149 & 1237 & $177.3846$ & $22.3967$ & $0.70$ & \Ha & $2.25$ & $1.36_{-0.00}^{+0.01}$ & $8.64\pm0.12$ & $15\pm36$ & $27\pm13$ & $4$ \\
MACS1149 & 1501 & $177.3970$ & $22.3960$ & $1.49$ & \OIII & $2.25$ & $12.42_{-1.43}^{+1.28}$ & $9.44\pm0.11$ & $15\pm7$ & $227\pm32$ & $2$ \\
MACS1149 & 1625 & $177.3900$ & $22.3895$ & $0.96$ & \Ha & $2.25$ & $1.80_{-0.02}^{+0.02}$ & $10.54\pm0.12$ & $123\pm5$ & $139\pm11$ & $4^f$ \\
MACS1149 & 1644 & $177.3944$ & $22.3892$ & $0.96$ & \Ha & $2.25$ & $1.79_{-0.02}^{+0.03}$ & $10.35_{-0.18}^{+0.12}$ & $200\pm12$ & $108\pm24$ & $3$ \\
MACS1149 & 1757 & $177.4085$ & $22.3868$ & $1.25$ & \OIII & $2.25$ & $3.83_{-0.11}^{+0.13}$ & $8.39_{-0.18}^{+0.12}$ & $5\pm29$ & $16\pm28$ & $2^e$ \\
MACS1149 & 1931 & $177.4034$ & $22.3816$ & $1.41$ & \OIII & $2.25$ & $2.05_{-0.03}^{+0.03}$ & $10.22\pm0.10$ & $44\pm4$ & $137\pm13$ & $2$ \\
MACS2129 & 37 & $322.3627$ & $-7.7099$ & $2.29$ & \OII & $1$ & $1.58_{-0.06}^{+0.07}$ & $10.13_{-0.15}^{+0.11}$ & $94\pm7$ & $-$ & $5$ \\
MACS2129 & 49 & $322.3528$ & $-7.7101$ & $1.88$ & \OII & $1$ & $1.38_{-0.02}^{+0.01}$ & $9.80_{-0.20}^{+0.14}$ & $109\pm45$ & $-$ & $5$ \\
MACS2129 & 329 & $322.3634$ & $-7.7032$ & $1.65$ & \OIII & $1$ & $1.67_{-0.05}^{+0.06}$ & $9.74_{-0.17}^{+0.12}$ & $11\pm6$ & $132\pm23$ & $2$ \\
MACS2129 & 1437 & $322.3508$ & $-7.6819$ & $1.36$ & \OIII & $1$ & $1.66_{-0.02}^{+0.02}$ & $9.52\pm0.10$ & $38\pm18$ & $100\pm50$ & $2$ \\
MACS2129 & 1566 & $322.3724$ & $-7.6792$ & $1.48$ & \OIII & $1$ & $1.56_{-0.05}^{+0.07}$ & $9.56\pm0.12$ & $101\pm18$ & $-$ & $5$ \\
MACS2129 & 1739 & $322.3605$ & $-7.6745$ & $1.49$ & \OIII & $1$ & $1.41_{-0.02}^{+0.02}$ & $9.32\pm0.10$ & $60\pm6$ & $52\pm9$ & $4^f$ \\
RXJ1347 & 188 & $206.8861$ & $-11.7338$ & $0.93$ & \Ha & $3$ & $2.42_{-0.04}^{+0.04}$ & $10.10\pm0.10$ & $6\pm1$ & $87\pm12$ & $2$ \\
RXJ1347 & 287 & $206.8910$ & $-11.7474$ & $1.01$ & \OIII & $3$ & $2.34_{-0.07}^{+0.08}$ & $9.58\pm0.11$ & $39\pm1$ & $46\pm6$ & $2$ \\
RXJ1347 & 450 & $206.8757$ & $-11.7645$ & $0.85$ & \Ha & $3$ & $1.92_{-0.03}^{+0.03}$ & $10.12\pm0.10$ & $37\pm54$ & $145\pm50$ & $2^e$ \\
RXJ1347 & 472 & $206.8719$ & $-11.7610$ & $0.91$ & \Ha & $3$ & $2.78_{-0.07}^{+0.10}$ & $9.44_{-0.23}^{+0.15}$ & $42\pm3$ & $108\pm8$ & $1$ \\
RXJ1347 & 795 & $206.8882$ & $-11.7613$ & $0.62$ & \Ha & $3$ & $1.51_{-0.03}^{+0.02}$ & $10.33\pm0.10$ & $45\pm2$ & $198\pm8$ & $2$ \\
RXJ1347 & 1230 & $206.8960$ & $-11.7537$ & $1.77$ & \OIII & $3$ & $40.56_{-12.83}^{+32.59}$ & $8.76_{-0.40}^{+0.21}$ & $62\pm19$ & $172\pm49$ & $2^e$ \\
RXJ1347 & 1261 & $206.9002$ & $-11.7476$ & $0.61$ & \Ha & $3$ & $1.76_{-0.03}^{+0.02}$ & $10.47\pm0.10$ & $12\pm6$ & $217\pm17$ & $1$ \\
RXJ1347 & 1419 & $206.9022$ & $-11.7443$ & $1.14$ & \OIII & $3$ & $8.98_{-0.60}^{+0.80}$ & $9.30\pm0.11$ & $45\pm2$ & $25\pm8$ & $2$ \\

\hline
\multicolumn{12}{l}{\textsc{Note.} -- $^a$ IDs match the GLASS IDs in the public data release, v001 available at \url{https://archive.stsci.edu/prepds/glass/}.}\\
\multicolumn{12}{l}{$^b$ Magnification estimate from cluster mass maps as described in Section~\ref{sec:res_phot}.}\\
\multicolumn{12}{l}{$^c$ Stellar masses obtained from SED fitting (Section~\ref{sec:res_phot}) have been corrected for magnification.}\\
\multicolumn{12}{l}{$^d$ Kinematic class described in Section~\ref{sec:res_kinematics}. $^e$ Modeled with 2D velocity map model rather than \galpak. $^f$ Probable mergers.}
\end{tabular}
\end{table*}

\subsection{Photometric properties and gravitational lens modeling}\label{sec:res_phot}

The stellar masses are obtained from the spectral energy distributions (SEDs) of the galaxies in the HFF photometric catalogs \citep[MACS0416 and MACS1149,][]{Morishita2016}, and the CLASH photometric catalog \citep[MACS2129 and RXJ1347,][]{Postman2011}. The SEDs are fit using the Fitting and Assessment of Synthetic Templates (FAST) code \citep{Kriek2009} using the \citet{Bruzual2003} stellar populations with an exponential declining star formation history (SFH) and a \citet{Chabrier2003} initial mass function (IMF). Reddening for the stellar continua, $A_\mathrm{V,SED}$ are obtained for a \citet{Calzetti2000} dust extinction law.

\begin{figure}[t!] 
\includegraphics[width=0.45\textwidth]{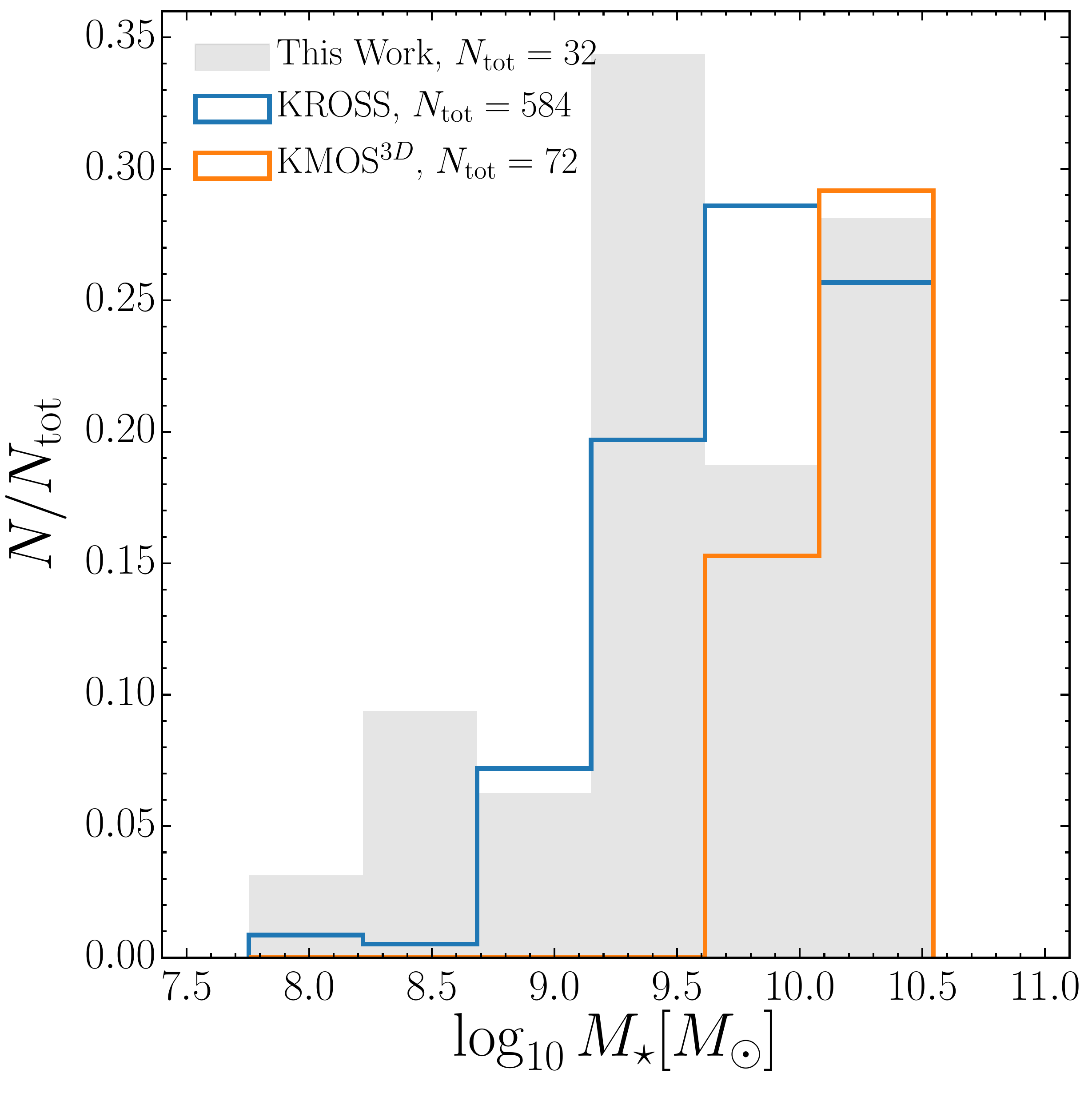}
\caption{Histograms showing the normalized stellar mass distributions of our sample compared to those of KROSS \citep[][$z\sim1$]{Stott2016} and \kmosd \citep[][$z\sim1$ and $z\sim2$]{Wisnioski2015}. A high proportion of the KLASS sample ($\sim19\%$) contains galaxies with stellar mass below $10^9M_\odot$, compared to $\sim5\%$ in KROSS, and no objects in \kmosd. While KROSS contains objects at $z\sim1$ with comparable stellar masses to KLASS, KLASS is the first survey to resolve kinematics in galaxies with stellar mass below $\sim6\times10^8 M_\odot$ at $z>1$.}
\label{fig:hist_mass}
\end{figure}

In Figure~\ref{fig:hist_mass} we show the demagnified stellar mass distribution of objects in our sample, compared with the \kmosd and KROSS samples. While we have a smaller sample than those surveys, we cover a broader mass range and have larger proportion of low stellar mass objects. $\sim63\%$ of our sample is comprised of objects with stellar mass below the \kmosd mass limit of $\sim6\times10^9 M_\odot$. KROSS does contain objects with comparable stellar masses to KLASS, but in lower proportions: $\sim5\%$ of the KROSS sample contains galaxies with stellar mass below $10^9 M_\odot$, compared to $\sim19\%$ in KLASS. KLASS is the first IFU survey to resolve kinematics in objects with stellar mass below $\sim6.3\times10^8 M_\odot$ at $z>1$: 5 objects in our sample fall below this mass limit.

The lensing effects on each source due to the gravitational potential of the foreground cluster were obtained using the SWUnited cluster mass models \citep{Bradac2005,Bradac2009}. For the clusters which are part of the HFF we use the publicly available lens models\footnote{\url{https://archive.stsci.edu/prepds/frontier/lensmodels/}} \citep{Hoag2016}, for MACS2129 we use the model described in \citet{Huang2016a}, and the modeling for RXJ1347 will be presented in Finney et al. in prep. Lensing parameters were obtained at the position of each targets from the 2D maps of $\kappa$, the dimensionless mass surface density of the lensing system, or \textit{convergence}, and $\gamma = \gamma_1 + i \gamma_2$ the \textit{shear} which measures the distortion of images \citep[e.g.,][]{Kochanek2004,Keeton2005}. Stellar masses and SFRs obtained from SED fitting and emission line flux measurements will be affected by gravitationally lensing, as fluxes are magnified. In the following sections we will refer to the magnification-corrected stellar masses and SFRs. Additionally, gravitational lensing distorts the images and kinematic maps, and removes symmetries in the velocity maps \citep{DeBurgh-Day2015}.

In this first exploration of the sample, we chose not to do a full lensing reconstruction of the objects back to the source plane. The majority of the sources in this paper are not significantly lensed, with magnifications $\mu \sim 1.4-2.5$, and have small shears $|\gamma| \simlt 0.1$, so the effect of lensing is small on the kinematic maps.

However, gravitational lensing will distort the axis ratios of objects derive from photometry, which are needed for deriving effective radii and inclination angles, so when using these values we scale by the extra distortion induced by lensing. Following \citet{Keeton2001}, we derive the source plane axis ratio from the observed image plane axis ratio, $(b/a)_\mathrm{im}$ as:
\BEA \label{eqn:cosi}
\left(\frac{b}{a}\right)_\mathrm{s} &=& \frac{1}{q_L} \left(\frac{b}{a}\right)_\mathrm{im}, \\
\mathrm{where}\; q_L &=& \frac{1 - \kappa + |\gamma|}{1 - \kappa -|\gamma|}
\EEA
is the inverse of the axis ratio produced by lensing.

The source plane axis ratio is equal to $\cos{i}$ where ${i}$ is the intrinsic inclination angle of a thin disk. We note that the axis ratio depends on the position angle of the source from the center of the lensing potential. We assume an average source position angle of 90\degree to derive this equation. The source plane effective radius is approximated as $r_{e,s} = (1 - \kappa -|\gamma|)r_{e,\mathrm{im}}$. For the majority of sources presented here, the shear effects produce $<40\%$ changes in effective radii and axis ratios. In the cases where objects have $\mu\simgt4$ lensing effects on kinematics may be stronger, so we clearly mark these outliers on figures which follow.

KLASS is complemented by additional VLT follow-up of gravitationally lensed multiple images in the Frontier Fields clusters MACS0416 \citep[with VIMOS and MUSE,][]{Grillo2015,Caminha2016} and MACS1149 \citep[with MUSE,][]{Grillo2016}. There is excellent agreement in spectroscopic redshifts for overlapping objects in KLASS and these studies. MACS0416\_372 and MACS0416\_955 are multiple images of the same galaxy \citep{Grillo2015,Caminha2016}. The derived intrinsic velocity dispersions (see Section~\ref{sec:res_kinematics}) for these images from our data are consistent with being the same source (see Table~\ref{tab:targets}), however, their stellar masses differ by 0.43 dex. We note that stellar population parameters which are independent of magnification (age and specific SFR) derived from the SEDs of these objects are consistent, so this discrepancy is likely due to systematics in the magnification map we used. Thus this is a good example of the complexities of cluster mass modeling and the need for many spectroscopically confirmed multiple images. MACS0416\_394 is also a multiple image, which may be overlapping with an image of the same system \citep{Caminha2016}, we thus neglect it when fitting any trends to our data. MACS1149\_1501 and MACS1149\_862 are multiple images of the SN Refsdal host galaxy \citep{Kelly2015,Treu2016,Grillo2016}.

\subsection{Kinematics}\label{sec:res_kinematics}

We derive kinematic properties of the galaxies by fitting the strongest available emission line: \Ha at $z < 1$, the \OIII doublet at $1 \leq z < 1.8$, and the \OII doublet at $z\geq1.8$. 

To measure integrated properties, we sum the flux in spaxels within a spatial aperture derived from the whitelight image of the full cube, which is then expanded to optimize the signal-to-noise ratio in the line. This results in a custom-extracted 1D spectrum for each galaxy. We then fit the emission lines as a Gaussian (two Gaussians for the doublets), plus a linear continuum component. The fits are weighted by the inverse variance spectrum derived from the sky emission lines and the bad pixel mask. To measure the total velocity dispersion in the ionized gas we subtract the instrumental broadening (FWHM $\sim 4\AA$) in quadrature and to correct for beam smearing due to the PSF we subtract the radial velocity gradient linearly \citep{Stott2016}:

\BE \label{eqn:sigma0}
\sigma_0^2 = \left(\sigma_\textrm{obs} - \frac{\Delta V}{\Delta R_\textsc{psf}}\right)^2 - \sigma^2_\textrm{instr}
\EE
where $\sigma_\textrm{obs}$ is the observed velocity dispersion measured from the 1D emission line fitting, $\Delta V/\Delta R_\textsc{psf}$ is the velocity gradient measured within the PSF radius, and $\sigma_\textrm{instr}$ is the instrumental broadening measured from sky lines in the data. If no clear velocity gradient is measured we subtract 23.3 km/s, the median beam smearing value for the sample. We note that beam smearing is highly dependent on the flux profile of the individual galaxy compared to the size of the PSF, so this median value may not be suitable in all cases, but it provides a reduction of an appropriate order.

Kinematic maps are produced by fitting Gaussian profiles to the flux in individual spaxels over a 50\AA~range around the linecenter derived from the integrated emission line profile. We require a S/N $>5$ for a successful fit. Following~\citet{Livermore2015,Stott2016} we initially aim to fit the emission line in 1 spaxel ($0\farcs2 \times 0\farcs2$), but expand the fitting region to encompass neighboring spaxels in $0\farcs4 \times 0\farcs4$ and $0\farcs6 \times 0\farcs6$ apertures if the S/N criterion is not satisfied. We then reject any spaxels where the errors on the derived flux, central wavelength or line dispersion are $>50\%$ of the measured values. The velocity and velocity dispersion maps are produced as the 1st and 2nd moments of the line profiles. The high S/N 2D flux and kinematic maps for our sample are presented in Appendix~\ref{app:maps}. We are able to produce kinematic maps for 25/32 galaxies.

To derive kinematic properties we use the kinematic fitting code \galpak \citep{Bouche2015} which was designed for IFU instruments and fits disk models to 3D data cubes themselves \citep[see also][for an alternative 3D fitting tool]{DiTeodoro2015,DiTeodoro2016}. \galpak has been shown to work well with VLT/Multi Unit Spectroscopic Explorer (MUSE) data \citep{Bacon2015,Contini2016}. This is the first time it has been used on KMOS data. We refer the reader to \citet{Bouche2015} for a full description of the fitting procedure, and we outline only the key points here.

Using \galpak we produce a three-dimensional galaxy model with an exponential radial flux profile and Gaussian disk luminosity thickness with scale height $h_z$. The velocity profile is modeled as an arctangent function~\citep{Courteau1997a}:

\BE \label{eqn:arctan}
V(r) = V_\textrm{max} \sin{i} \frac{2}{\pi} \arctan{\left(\frac{r}{r_t}\right)}
\EE

where $r$ is the distance along the major axis of the galaxy in the plane of the sky, $V_\textrm{max}$ is the asymptotic velocity in the plane of the disk, $i$ is the inclination of the disk, and $r_t$ is the turnover radius. 

The total line-of-sight velocity dispersion, $\sigma_\textrm{tot}$ has three components: (1) the local isotropic velocity dispersion, $\sigma_d$, due to the self-gravity of the disk, which is given by $\sigma_d(r)= h_z V(r)/r$ for a compact thick disk; (2) a mixing term due to the mixing of velocities along the line of sight in a disk with non-zero thickness; and (3) the intrinsic velocity dispersion, $\sigma_0$, which we assume to be isotropic and spatially constant, which measures the dynamical `hotness' of the disk. The terms are added in quadrature:

\BE \label{eqn:veldisp}
\sigma_\textrm{tot}^2(r) = \sigma_d(r)^2 + \sigma_m^2 + \sigma_0^2
\EE

By construction, $\sigma_\textrm{tot}$ is radially symmetric. It is therefore likely that the galaxies best fit by \galpak have the most regular symmetric velocity dispersion maps. However, the observed total velocity dispersion maps of many galaxies in our sample, plotted in Figures~\ref{fig:postage1}-\ref{fig:postage5}, are not symmetric. This asymmetry may be due to clumpy star formation and/or slow mixing of turbulence in the disk \citep{Genzel2011,Glazebrook2013}. 

The model is convolved with a 3D kernel comprised of the instrumental line spread function (LSF) of KMOS measured from sky lines and the median PSF of the data derived from observing faint stars (see Section~\ref{sec:data_design}). This 3D convolved model is then compared with the data itself. We use the Bayesian MCMC fitting method described in \citet{Bouche2015} to fit the 3D model to the data cubes in the vicinity of the emission line. We fix the $x, y$ centroid of the emission line to the center of the YJ continuum and fit for 9 free parameters: $\lambda_c$ the central wavelength of the emission line; $f_\textrm{tot}$ the total flux in the emission line; $r_{1/2}$, the half-light radius of the disk; the inclination $i$ of the disk and the position angle (PA) in the plane of the sky; the velocity profile turnover radius $r_t$; the asymptotic velocity $V_\textrm{max}$; a systemic velocity offset $V_\textrm{sys}$; and the intrinsic velocity dispersion $\sigma_0$. As inclination, $r_t$ and $V_\textrm{max}$ are degenerate in an arctangent model, if a source is not highly distorted by lensing magnification ($\mu < 4$) we additionally constrain the inclination using a uniform prior over $i_\mathrm{HST}\pm20$\degree, where $i_\mathrm{HST}$ is obtained from the axis ratio of sources in the HST photometry (Section~\ref{sec:res_phot}) using GALFIT \citep{Peng2010} and assuming a thin disk.

Using a Bayesian method enables us to treat the uncertainties on all parameters simulatenously and robustly. We require the acceptance rate of useful iterations of the MCMC walk to be between $30-50\%$, the reduced$-\chi^2$ of the model velocity map to be $<15$, and the RMS difference between the observed and model velocity maps to be $<0.4 V_\textrm{max} \sin{i}$ for fit to be accepted. 11 objects are fit by \galpak.

Some objects display clear velocity gradients in the 2D kinematic maps (Figures~\ref{fig:postage1}-\ref{fig:postage5}) but \galpak fits them poorly. For these objects we fit the 2D velocity map using Equation~\ref{eqn:arctan}, rather than the full 3D fitting. We fit the 2D maps using a Bayesian model to fit for: the inclination $i$ of the disk and the position angle (PA) in the plane of the sky; the velocity profile turnover radius $r_t$; the asymptotic velocity $V_\textrm{max}$; and a systemic velocity $V_\textrm{sys}$. We use the \textsc{EMCEE} MCMC sampler \citep{Foreman-Mackey2013} and the same prior on inclination as described above. This 2D method produced kinematic parameters consistent with those from \galpak for the same objects. We accept 5 objects for which \galpak failed as well-fit by the 2D model, with the reduced$-\chi^2$ of the fit $<15$, and the RMS difference between the observed and model velocity maps to be $<0.4 V_\textrm{max} \sin{i}$.

Of these 5 objects which were fit only by the 2D model, 2 are significantly magnified, with $\mu>15$ (MACS0416\_394 and RXJ1347\_1230). As \galpak attempts to fit a model to the light profile of the galaxy, it is expected that \galpak will fail for highly magnified objects which are distorted. As lensing doesn't change the observed velocities \citep{DeBurgh-Day2015,Jones2010a} it is still possible to fit a simple kinematic model to these objects, but due to the large uncertainties in measuring inclination deriving source plane parameters is challenging (see Section~\ref{sec:res_phot} for more discussion), so these objects are neglected in our further analyses. The other 3 objects (MACS0416\_863, MACS1149\_1757 and RXJ1347\_450) are compact, with $r_{1/2} < r_\textrm{PSF} \sim 0\farcs3$, so were unlikely to be well-fit by \galpak, which requires $r_{1/2} \simgt 1.5 r_\textrm{PSF}$.

To transform to the source plane, we correct the inclination and radii fit by \galpak or the 2D method using Equation~\ref{eqn:cosi}, and then calculate a source plane $V_\textrm{max}$ using the corrected inclination.

For the remaining objects which were not well-fit by either \galpak or the 2D velocity map model we construct $V_\textrm{max,obs} = (v_\textrm{max} - v_\textrm{min})/2$ from the 2D velocity maps as an approximate measure of the rotational velocity of these objects, which are unlikely to be regularly rotating. This measure of $V_\textrm{max}$ is likely to underestimate an asymptotic velocity as we do not measure velocity with high S/N in the outer regions of many galaxies. Thus we rescale velocities measured in this way, assuming arctangent rotation curves (Equation~\ref{eqn:arctan}):

\BE \label{eqn:vscale}
V_\textrm{max} = \frac{V_\textrm{max,obs}}{\sin{i}} \frac{\pi}{2}\frac{1}{\arctan{(R/\bar{r_t})}}
\EE

where $i$ is the source plane inclination determined from photometry (Section~\ref{sec:res_phot}) and corrected for lensing via Equation~\ref{eqn:cosi}), $R$ is a measure of the observed radial extent of the galaxy in KMOS ($R = \sqrt{(N_\textrm{px}/\pi)}$, where $N_\textrm{px}$ is the number of spaxels in the 2D velocity map) and $\bar{r_t} = 1.24$ is the median turnover radius of the sample fit by \galpak and the 2D method described above. This results in a median rescaling factor for the velocities of these objects of 1.25.

These 3 measures of maximum velocity (from \galpak, the 2D fit, and the rescaled velocity map) produce consistent velocities for the 11 galaxies which were well-fit by \galpak, it is reasonable to use `last-resort' methods to estimate the velocity of rotating disks. We use velocities derived from \galpak and the 2D fit to investigate trends for objects which are likely rotating disks. The objects which were not fit by either \galpak or the 2D method do not have clear velocity gradients are unlikely to be rotating disks, here $V_\textrm{max}$ gives us an approximate measure of gas kinematics in the galaxies.

Following the fitting we classify the galaxies into 5 kinematic categories: 
\begin{enumerate}
\item Regular rotators: rotation dominated systems with $V_\textrm{max}/\sigma_0 > 1$ and well-fit by\galpak with the reduced$-\chi^2$ of the model velocity map $<3$ and the RMS difference between the observed and model velocity maps $<0.15 V_\textrm{max} \sin{i}$. These comprise 3/25 of the resolved systems.
\item Irregular rotators: systems with clear velocity gradients fit either by \galpak or the 2D method with reduced$-\chi^2$ of the fit $<15$ and the RMS difference between the observed and model velocity maps $<0.4 V_\textrm{max} \sin{i}$, and $V_\textrm{max}/\sigma_0 > 1$. These comprise 13/25 resolved systems.
\item Dispersion dominated: systems with $V_\textrm{max}/\sigma_0 < 1$. These comprise 3/25 resolved systems.
\item Mergers/unknown: systems with merging signatures evident in \HST images and/or kinematic maps, and systems with irregular kinematic maps where no clear velocity gradients are evident. These comprise 6/25 resolved systems, of which 2 appear to be mergers.
\item Unresolved: there are 7/32 total systems with unresolved kinematic maps. The majority of these systems are compact or in MACS2129, which had the shortest integration time. We measure the velocity dispersion of the these objects from their 1D spectra obtained in S/N optimized spatial apertures.
\end{enumerate}

Table~\ref{tab:targets} shows the sources presented in this paper and their derived kinematics properties. Observed kinematic maps for the sample are shown in Appendix~\ref{app:maps} in Figures~\ref{fig:postage1}-\ref{fig:postage5}. We plot the \HST RGB images obtained from CLASH or HFF photometry, the whitelight image from KMOS YJ, the 2D emission line used for kinematic modeling, the velocity maps and velocity dispersion maps.

Rotation curves are produced by plotting 1D cuts in the velocity maps, along the major kinematic axis of the galaxy as determined by the Bayesian fitting, either from \galpak or the 2D fitting. Rotation curves for the regular and irregular rotators (kinematic classes 1 and 2) are presented in Appendix \ref{app:rotcurves}.

The introduction of 2 classes of rotators was motivated by the 3 objects for which \galpak produced excellent fits, with the cut (reduced$-\chi^2$ of the model velocity map $<3$ and the RMS difference between the observed and model velocity maps $<0.15 V_\textrm{max} \sin{i}$ for class 1) arising from a natural clustering in the fit statistics for the rotating objects. The class 1 objects have symmetric rotation curves (Figure~\ref{fig:rotcurves_reg}) aligned with their photometric axes and velocity dispersion maps which appear to peak at the center of their light profiles (Figures~\ref{fig:postage1}, with the exception of MASC1149\_593). The second `irregularly rotating' group of galaxies display more asymmetries in their velocity and velocity dispersion maps (e.g. RXJ1347\_287 and RXJ1347\_188), and several have clumps in the HST photometry which may indicate intense star forming regions or minor mergers (e.g. MACS1149\_1757, RXJ1347\_1419 and RXJ1347\_795).

\subsection{Star formation drivers}\label{sec:res_SF}

\begin{figure}[t!] 
\includegraphics[width=0.49\textwidth]{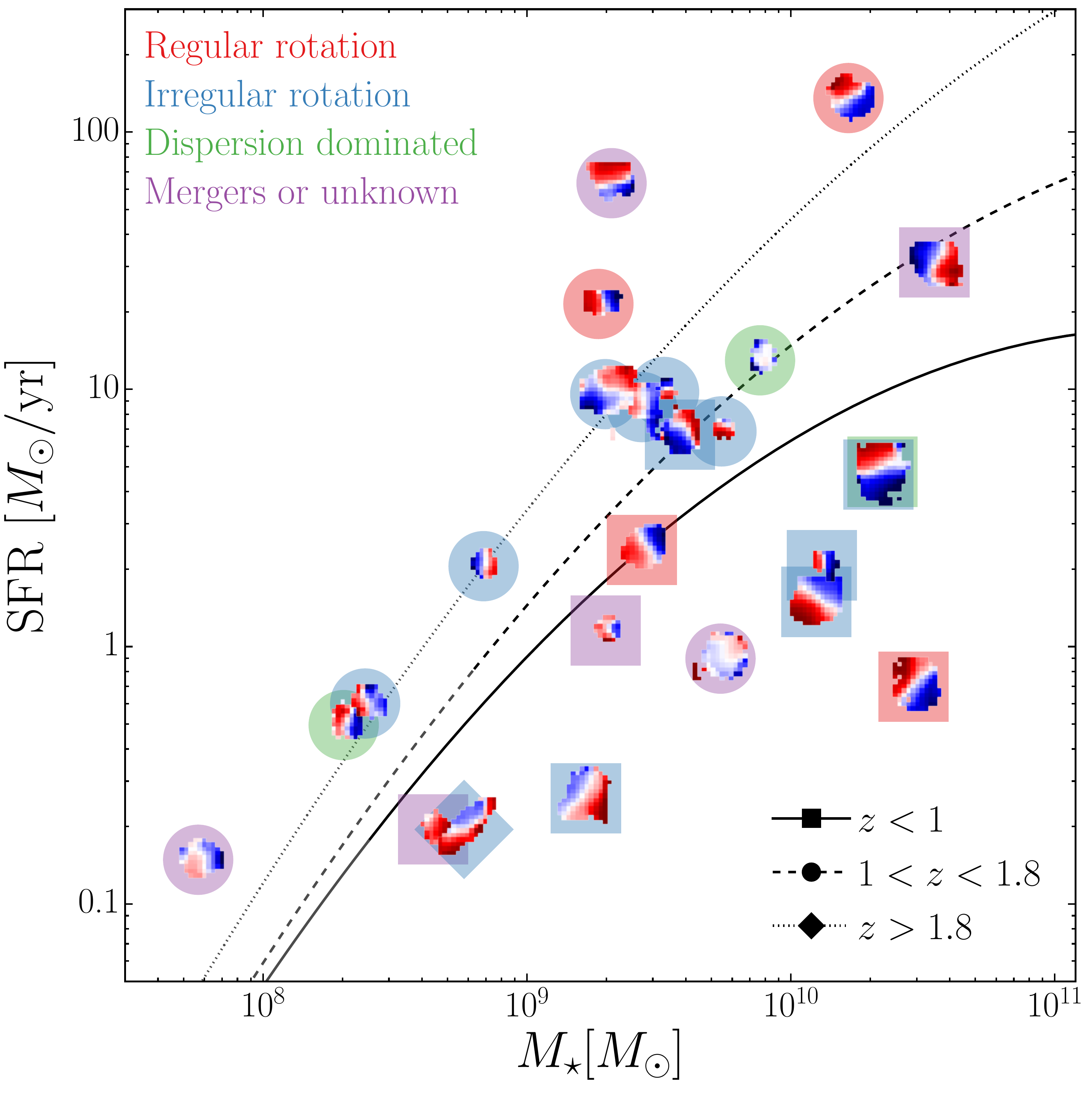}
\caption{The velocity maps for galaxies in our sample with resolved kinematics, plotted at the galaxy's position on the SFR$-\mstar$ plane. We plot the empirical SFR models from \citet{Whitaker2014} at a range of redshifts (indicated by linestyle) to compare with KLASS. The shapes around the velocity maps indicate the redshift bin of the object: $z<1$ (square), $1 \leq z < 1.8$ (circle) or $z\geq1.8$ (diamond), and the colors indicate the kinematic class as described in Section~\ref{sec:res_kinematics}. Our sample shows a large scatter around the star forming main sequence.}
\label{fig:sfr_mstar_velmaps}
\end{figure}

\begin{figure*}[t!]
\includegraphics[width=0.99\textwidth]{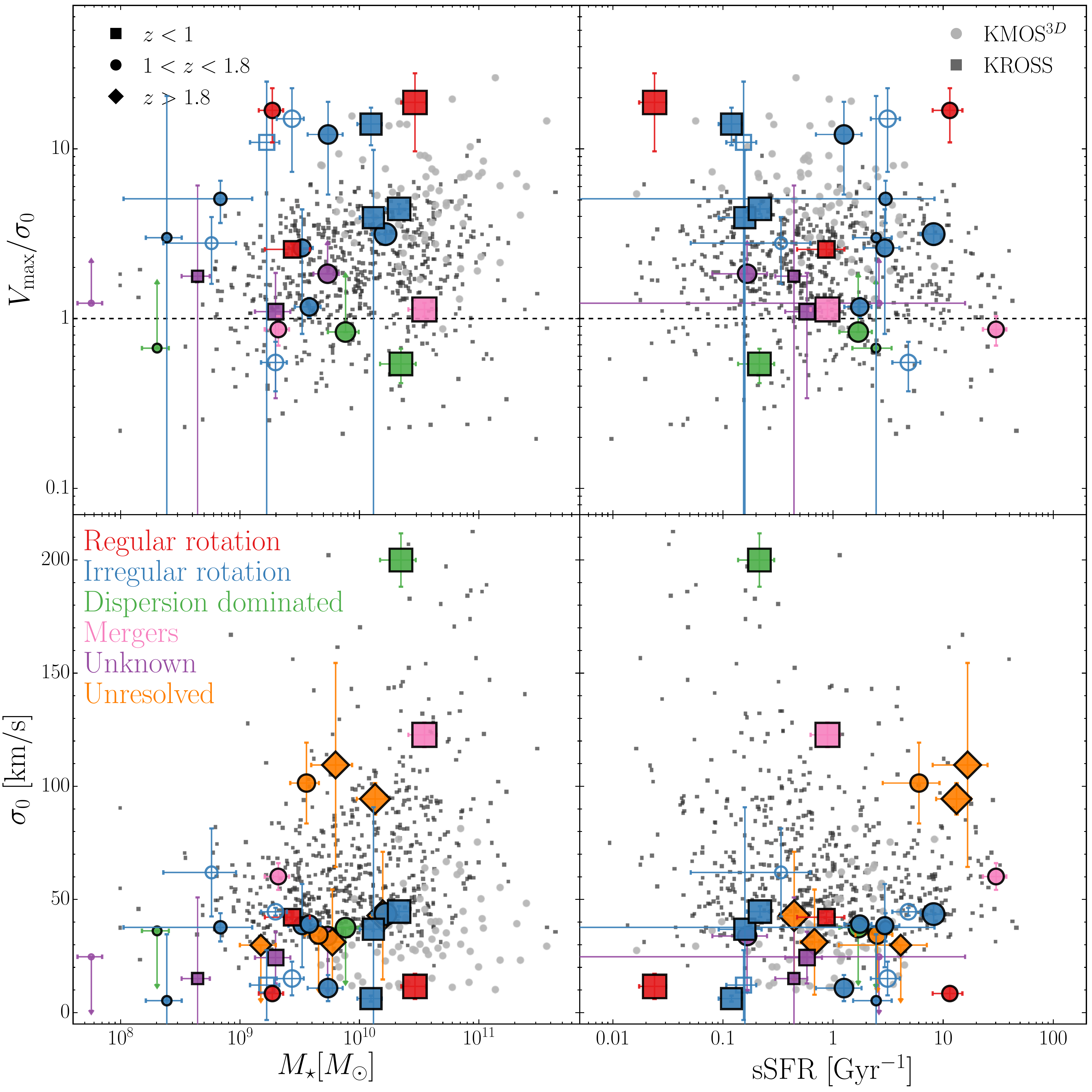}
\caption{Velocity dispersion and $V_\textrm{max}/\sigma_0$ as a function of stellar mass and sSFR for our sample. The marker styles indicate the redshift bins of the survey, the colors indicate the kinematic classification and the marker size indicates stellar mass. Open markers indicate the magnification of the object is large ($\mu > 4$) so intrinsic velocity measurements are more uncertain. We note that the majority of systems at $z>2$ which are unresolved are in the the cluster MACS2129 which had the shortest integration time. We plot points from KROSS \citep{Stott2016} and \kmosd \citep{Wisnioski2015} for comparison. We measure $\sigma_0$ in a similar way to KROSS (Section~\ref{sec:res_kinematics}), whereas \kmosd measuring $\sigma_0$ from the outer regions of 2D velocity dispersion maps. \citet{Stott2016} notes KROSS $\sigma_0$ obtained in this way are a factor $\sim2$ higher than those from \kmosd.}
\label{fig:sfcorr}
\end{figure*}

To investigate the relationships between the high SFR observed at $z\simgt 1$ \citep[e.g.,][]{Madau2014,Whitaker2014} we look for correlations between SFRs and the kinematic properties of our sample, and any evolution with redshift.

To derive SFRs we use the KMOS measured magnification-corrected line flux of \Ha ($z < 1$), \Hb ($1 \leq z < 1.8$) or \OII ($z\geq1.8$). We convert line flux to SFR using relations from \citet{Kennicutt1998} and divide SFRs by a factor of $1.7$ to convert from a \citet{Salpeter1955} to \citet{Chabrier2003} initial mass function (IMF). We assume a Case B recombination extinction-corrected Balmer decrement of $2.86$ for all \Hb measurements.

To correct for extinction we use the stellar reddening factors calculated from the SED fitting and additional empirical calibrations to convert continuum extinction to nebular emission line extinction. Objects MACS1149\_593 and MACS1149\_683 have nebular extinction values derived from GLASS data by \citet{Wang2016}. For the remaining sources, we use the calibration from \citet{Wuyts2013} where $A_\mathrm{V,gas} = A_\mathrm{V,SED} (1.9 - 0.15 A_\mathrm{V,SED})$. We use the \citet{Cardelli1989} reddening curve with $R_V = 3.1$.

In Figure~\ref{fig:sfr_mstar_velmaps} we plot the galaxy stellar masses derived from photometry (Section~\ref{sec:res_phot}) versus their SFRs derived from the KMOS line fluxes. We see that our galaxies are scattered around the star forming main sequence in their mean redshift bins. Our sample is too small to draw firm conclusions, but reflects a large diversity in the lives and dynamics of star forming galaxies at cosmic noon.

In Figure~\ref{fig:sfcorr} we plot correlations between $V_\textrm{max}/\sigma_0$, $\sigma_0$, stellar mass and specific SFR. We see that the majority of our systems have $V_\textrm{max}/\sigma_0 > 1$ indicating most of these systems are rotationally supported. 16/25 resolved systems are rotationally supported (kinematic classes 1 and 2), with 5 additional objects with $V_\textrm{max}/\sigma_0 > 1$, consistent with the 83\% of systems in KROSS \citep{Stott2016} at $z\sim1$, and 93\% of systems at $z\sim1$ and 73\% of systems at $z\sim2$ in \kmosd \citep{Wisnioski2015}.

Of the 5 objects have $V_\textrm{max}/\sigma_0 < 1$, we consider only 3 to be dispersion dominated. RXJ1347\_1419 has a clear rotation curve so is classed as an irregular rotator (Figure~\ref{fig:rotcurves_irreg1}), but is magnified by $\mu\sim9$ so there are large uncertainties in deriving the source plane maximum velocity. MACS2129\_1739 has multiple components (Figure~\ref{fig:postage5}) and a velocity gradient, suggesting this is a merger system.

\subsection{Kinematic trends}\label{sec:res_kinematic_trends}

\begin{figure}[ht!]
\includegraphics[width=0.45\textwidth]{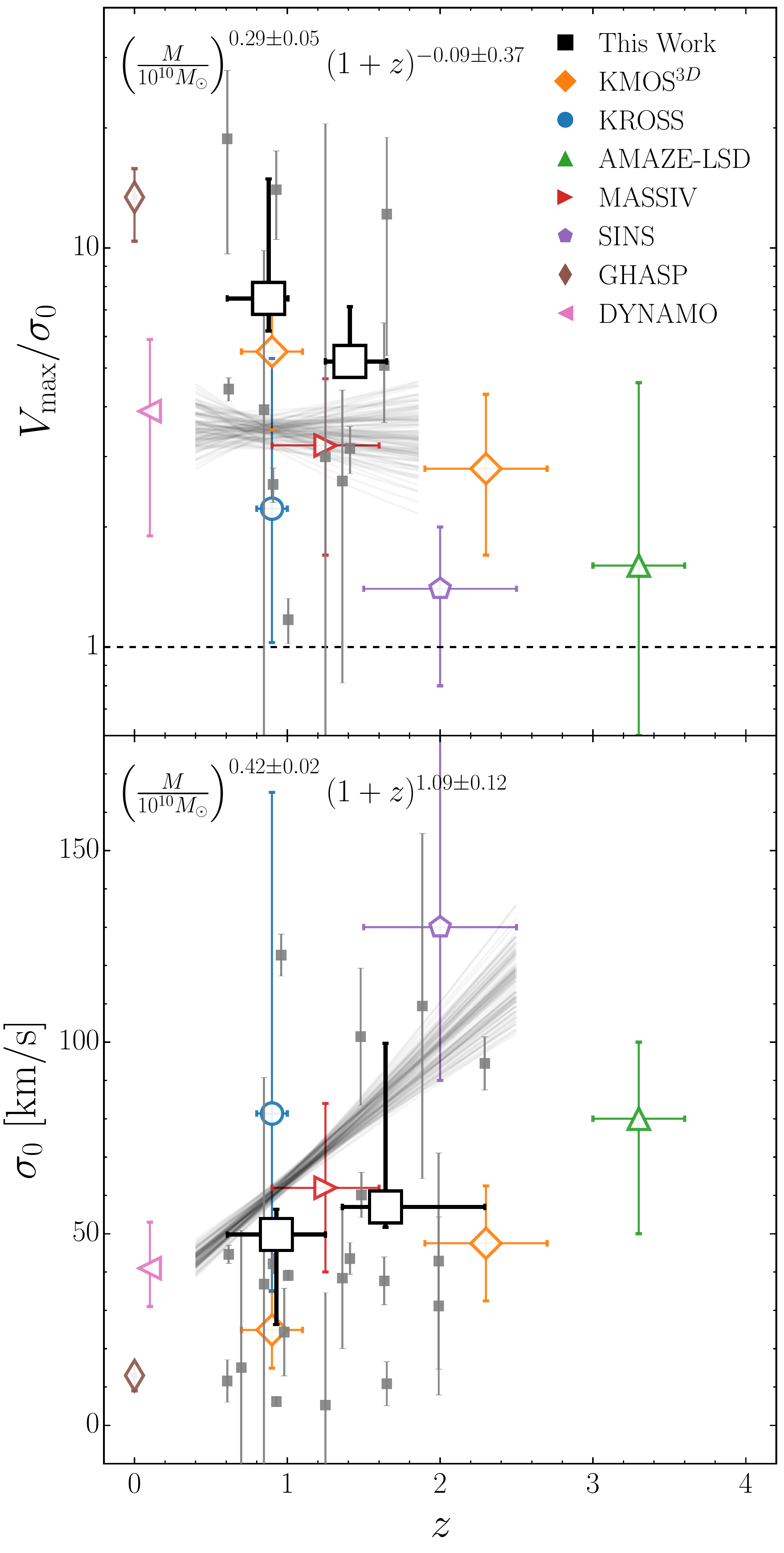}
\caption{$V_\textrm{max}/\sigma_0$ as a function of redshift for the galaxies classified as regular rotators and irregular rotators in our sample (Section~\ref{sec:res_kinematics}), and intrinsic velocity dispersion as a function of redshift for all galaxies in our sample. We exclude highly magnified galaxies ($\mu > 4$) and galaxies for which we have only limits on these parameters. We plot the individual data as gray squares with 1 standard deviation error bars, and our data in two redshift bins as black squares (the squares are positioned at the mean value within the redshift bin, the vertical bars show the 50\% range and the horizontal bars show the full redshift range of the bin. For comparison we plot the mean, 50\% value range (vertical bars) and redshift range (horizontal range) of rotation dominated galaxies in other IFU surveys: \kmosd \citep[orange diamonds,][]{Wisnioski2015}; KROSS \citep[blue circles,][]{Stott2016}; AMAZE-LSD \citep[green triangles,][]{Gnerucci2010}; MASSIV \citep[red right triangles,][]{Epinat2012}; SINS \citep[purple pentagons,][]{Schreiber2009}; GHASP \citep[brown triangles,][]{Epinat2008a,Epinat2008}; and DYNAMO \citep[pink left triangle,][]{Green2013}. We show draws from the MCMC samples for the redshift and mass dependent fit to our data as the gray lines (described in Section~\ref{sec:res_SF}) and give the fitted exponents for redshift and stellar mass evolution. The trend in $V_\textrm{max}/\sigma_0$ is dominated by stellar mass for our sample and an increase of velocity dispersion with increasing redshift is found for our sample.}
\label{fig:zevol}
\end{figure}

\begin{figure}[ht!]
\includegraphics[width=0.45\textwidth]{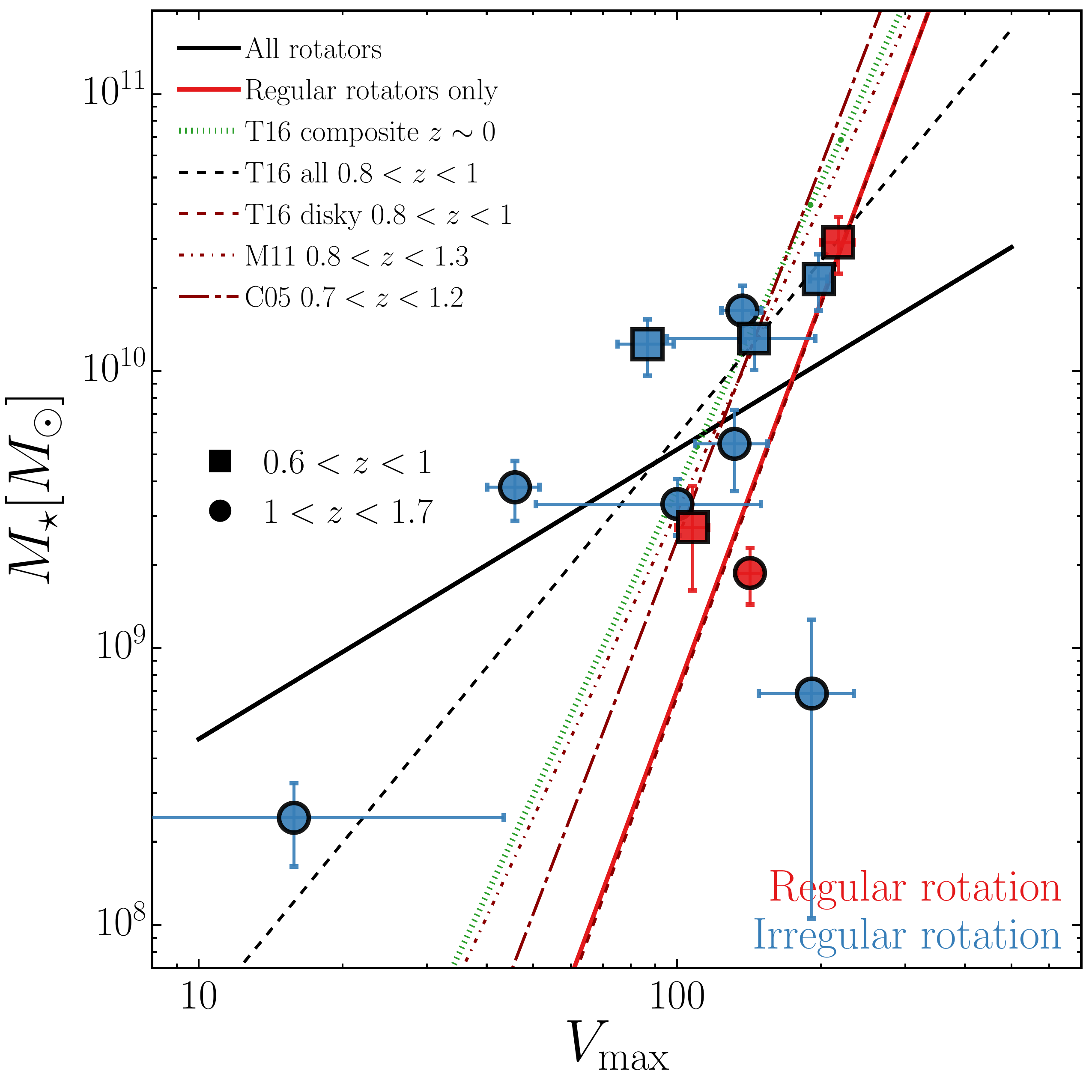}
\caption{Stellar mass Tully-Fisher relation with $V_\mathrm{max}$ for rotating galaxies in our sample (kinematic classes 1 and 2). As before, the colors indicate the kinematic classification. Objects with magnification $\mu > 4$ are excluded. For comparison we plot stellar mass Tully-Fisher relations from the literature over a similar redshift range \citep[red][]{Conselice2005,Miller2011,Tiley2016} and a $z\sim0$ relation (green dotted) from a compilation of data \citep{Pizagno2007,Rhee2004,Reyes2011} produced by \citet{Tiley2016}. From \citet{Tiley2016} we show the fits to their full rotator sample (dashed black) and their `disky' sample (dashed red) which is similar to our regular rotator sample. The best-fit relation to our full sample of rotators (solid black) is shallower than that of the regular rotators (solid red) and other relations from the literature. Our best-fit to the regular rotators is consistent with those from the literature.}
\label{fig:tf}
\end{figure}

KLASS shows a large diversity in the kinematics of star forming galaxies with mass and redshift. Regular rotating disks exist with stellar mass $>3\times10^{9}M_\odot$, with other irregular rotationally supported systems existing at a wider range of mass (upper left panel, Figure~\ref{fig:sfcorr}). The merging and irregular systems have sSFR $>0.1 \ssfrunit$ (right panels, Figure~\ref{fig:sfcorr}), suggesting that their disturbed gas dynamics may be enhancing the SFRs in some of these objects compared to kinematically ordered systems. Our sample shows higher sSFRs at higher redshift (lower right panel, Figure~\ref{fig:sfcorr}), irrespective of kinematic class, consistent with the expectation of a higher mass accretion rate and densities at high redshifts \citep{Tacchella2013,Mason2015a}.

With the power of lensing, we can access the low mass galaxies that are missing from other surveys.  Our sample includes 5 galaxies with stellar mass below $6.3\times10^8 M_\odot$, lower than any previous IFU targets at $z>1$. In Figure~\ref{fig:sfcorr} we can see that all the galaxies below $3\times10^{9}M_\odot$ have complex kinematics: no regular rotators exist below this limit. These systems at $z>1$ all have sSFR $>1\ssfrunit$. We discuss the low mass galaxies in Section~\ref{sec:dis_lowmass}.

The dispersion dominated systems exist at both low and high mass, and a decade apart in sSFR. These are clearly very different systems (see kinematic maps in Figure~\ref{fig:postage3}) and are discussed further in Section~\ref{sec:dis_dispersion}.

The mean $V_\textrm{max}/\sigma_0$ for the regular rotator sample is $12.7\pm2.4$ - similar to the values in the local universe \citep[$5-20$, e.g.][]{Epinat2010}, whilst for the irregularly rotator sample the mean value is lower at $5.5\pm0.5$, suggesting these objects are dynamically hotter. This justifies the splitting of the rotation dominated sample into 2 sub-samples.

In Figure~\ref{fig:zevol} we plot $V_\textrm{max}/\sigma_0$ for the rotation dominated galaxies in our sample, and $\sigma_0$ for all galaxies, as as function of redshift, to explore evolution in these parameters. We exclude 5 highly magnified galaxies ($\mu > 4$) and any galaxies for which we only have limits on these parameters. We also compare our data with results from other IFU surveys over the redshift range $0 \simlt z \simlt 3.5$. Comparing the data in redshift bins across all surveys, our data qualitatively support a trend of decreasing $V_\textrm{max}/\sigma_0$ with redshift \citep{Wisnioski2015,Glazebrook2013}, suggesting that disk galaxies are dynamically `hotter' at high redshift. 

To quantify any evolution in our sample we fit a simple model of the form $f \sim (\mstar/10^{10}M_\odot)^\alpha (1+z)^\beta$ using a Bayesian fitting procedure using the \textsc{EMCEE} sampler \citep{Foreman-Mackey2013}. We include dependence on mass because our sample spans a wide range in stellar mass which is correlated with velocity and, to some extent, velocity dispersion \citep[e.g.][]{Tully1977a,Kassin2007}. For $V_\textrm{max}/\sigma_0$, $\alpha=0.29\pm0.05$ and $\beta=-0.08\pm0.37$, suggesting a very marginal decline with increasing redshift, but no redshift evolution is also consistent with the data. There is a stronger dependence on stellar mass, which is expected because the highest stellar mass objects are predominantly regular rotating disks, as discussed above. For $\sigma_0$, $\alpha=0.42\pm0.02$ and $\beta=1.09\pm0.12$, indicating an increase in $\sigma_0$ with both increasing stellar mass and redshift for our sample. An increase in $\sigma_0$ with redshift is also seen in other work \citep[][]{Wisnioski2015} and could be due to high densities and high rates of gas inflow at high redshift driving up velocity dispersion in a disk.

We investigate the importance of rotational velocity in supporting the sample in Figure~\ref{fig:tf}. We plot the stellar mass Tully-Fisher relation between $\mstar$ and $V_\textrm{max}$. We plot only the galaxies classified as rotating (classes 1 and 2) and exclude 4 high magnification objects due to the uncertainties in deriving source plane velocities. For comparison, we plot best-fit relations from the literature at similar redshifts \citep{Conselice2005,Miller2011,Tiley2016} and a $z\sim0$ relation from a compilation of data \citep{Pizagno2007,Rhee2004,Reyes2011} produced by \citet{Tiley2016}. We see that the regularly rotating systems are closest to the fiducial Tully-Fisher relation, whilst the irregularly rotating systems are mostly scattered below the relation. Whilst this is a very small sample, it is consistent with the conclusions of \citet{Tiley2016}, which also found a large number of their rotating sample offset below the $z\sim0$ relation, but present a `disky' sub-sample (similar to our regular rotators) which lies closer to the $z\sim0$ relation. Most previous studies on the Tully-Fisher relation at $z\sim1$ \citep[see also][]{DiTeodoro2016} have preselected galaxies with disk morphologies, and thus are likely to miss objects such as those in our irregularly rotating sample which may not look like disks in photometric surveys but have velocity gradients.

We fit a linear relation ($\log_{10}{M_\star} = m \log_{10}{(V/V^*)} + c$, where $V^*$ is the median velocity in the sample) with intrinsic scatter $\sigma$, using a Bayesian technique with EMCEE \citep{Foreman-Mackey2013}, to the total sample of 16 objects and to the 3 regular rotators only. For the total sample: $\log_{10} V^*=2.13$, $m=1.03^{+0.52}_{-0.39}$, $c = 9.85^{+0.14}_{-0.15}$ and $\sigma = 0.43^{+0.15}_{-0.10}$. For the regular rotators only: $\log_{10} V^*=2.14$, $m = 4.66^{+1.62}_{-1.38}$, $c = 9.56^{+0.19}_{-0.17}$ and $\sigma = 0.05^{+0.59}_{-0.05}$. The regular rotator sample best-fit relation is consistent with the plotted literature relations, but with large uncertainty due to the small sample size. The full sample is consistent with \citet{Tiley2016} and suggests an additional source of support other than rotational velocity in the irregularly rotating galaxies.

Higher velocity dispersion may provide increased pressure support in disks, to investigate this we attempt to fit a similar relation between stellar mass and $S_{0.5} = \sqrt{V^2_\textrm{max}/2 + \sigma^2_0}$ as introduced by \citet{Kassin2007} to the sample in Figure~\ref{fig:tf}. \citet{Kassin2007} found reduced intrinsic scatter when including velocity dispersion. For the total sample our best-fit relation is $\log_{10} S_{0.5}^*=1.99$, $m = 1.19^{+0.65}_{-0.47}$, $c = 9.82^{+0.14}_{-0.15}$ and $\sigma = 0.42^{+0.15}_{-0.10}$. There is no significant decrease in intrinsic scatter between the pure velocity and $S_{0.5}$ Tully-Fisher relations for our sample. Thus it is unclear that intrinsic velocity dispersion provides significant pressure support to the systems presented here.

We urge caution before over-interpreting these plots which are influenced strongly by the selection sample and other potential biases of our small sample, including uncertainties in the magnification models. It is clear that there is much observational work to be done to build large representative samples to further investigate redshift trends of kinematic properties.

\section{Discussion}\label{sec:discussion}

\subsection{What is the dynamical nature of galaxies at $z \simgt 1$?}\label{sec:dis_class}

Integral fields surveys \citep{ForsterSchreiber2006,Genzel2011,Gnerucci2010,Flores2006} of $z\sim1-3$ galaxies have found rotation dominated systems, dispersion dominated systems, and merging/irregular systems, in roughly equal proportions. High redshifts disks were expected to be highly turbulent as rotation dominated systems had systematically higher velocity dispersions than local disks \citep{Epinat2010,Bershady2010}. This is in contrast to the local universe where most objects with stellar mass over $10^{10} M_\odot$ are dispersion dominated ellipticals, the highest mass objects at $z\simgt1$ were rotating disks.

Our sample shows a large diversity in the kinematic nature of galaxies at $z\simgt1$. The higher spatial resolution of KLASS compared to field surveys, due to the boost from lensing, has enabled us to clearly resolve rotating objects in our sample. The majority of our sample is rotation supported (16/25), but we find justification to define two rotation supported sub-samples: (1) regular rotators (3 objects) which are kinematically regular, with mean $V_\textrm{max}/\sigma_0 = 12.7\pm2.4$, similar to values for local disks \citep{Epinat2010}, and (2) irregular rotators with more disturbed kinematics and lower mean $V_\textrm{max}/\sigma_0 = 5.5\pm0.5$, suggesting dynamically hotter disks. This small fraction of galaxies exhibiting regular rotation was also seen recently in \citet{Leethochawalit2016a} using AO observations of lensed galaxies. Galaxies at high redshift are likely ongoing morphological and kinematic changes before settling into the bimodality we see in the local universe.

\subsection{Revealing the kinematics of low mass galaxies}\label{sec:dis_lowmass}

Gravitational lensing gives us access to the low mass galaxies missing from other surveys: KLASS has resolved kinematics in 5 galaxies at $z>1$ with stellar mass below $6.3\times10^8 M_\odot$, lower than any previously studied.

From the kinematic maps (Figures~\ref{fig:postage1}-\ref{fig:postage5}) and Figure~\ref{fig:sfcorr} we see that none of these low mass galaxies are regularly rotating: 2 are irregular rotators (MACS1149\_1757 and RXJ1347\_1230), 1 is dispersion dominated (MACS0416\_706), and 2 have unknown/merging kinematic structure (MACS1149\_683 and MACS1149\_1237). In Figure~\ref{fig:sfcorr} (right panels) we show that these systems also have sSFR $>1\ssfrunit$. Our data suggest the turbulent nature of star formation in low mass galaxies effects the kinematics of the whole galaxy: these low mass galaxies are all kinematically disturbed and rapidly star forming.

\subsection{What are dispersion dominated galaxies?}\label{sec:dis_dispersion}
High redshift dispersion dominated systems were first seen by \citet{Erb2006}. This population has been observed at stellar mass ranges $\sim1-5 \times 10^{10}M_\odot$ (the majority of high mass objects are observed with dominant rotation), and the number density of the population increases with redshift \citep{Law2007,Law2009,Epinat2012,Newman2013a}. The formation and evolution history of these objects is a relative mystery: perhaps formed by the collapse of a single molecular cloud; will they grow to large elliptical galaxies at low redshift, via mergers, or will they fade to remain relatively low mass?

High spatial resolution is needed to clearly distinguish rotationally supported galaxies from mergers and pressure supported systems. AO on Keck/OSIRIS and VLT/SINFONI have enabled high spatial resolution spectroscopy ($0\farcs2 \sim 1.7$ kpc at $z\sim1$) of a handful of $z\sim1-3$ galaxies, including objects which are gravitationally lensed and an order of magnitude lower in stellar mass than unlensed objects \citep{Jones2010a,Livermore2015}. These studies indicated that high fractions of rotating galaxies would be misclassified as dispersion dominated at seeing-limited resolution. This was confirmed with AO followed of SINS galaxies: \citet{Newman2013a} found the fraction of dispersion dominated systems in their sample dropped from 41\% to $6-9\%$ when these galaxies were observed with AO. Additionally, surveys with AO \citep{Gnerucci2010} and lensing \citep{Leethochawalit2016a} have also shown that galaxies classified as rotators in seeing limited conditions have irregular velocity maps when observed with higher spatial resolution. 

It is likely that the population of rotation dominated systems may be overestimated by low-resolution spectroscopy: most of these objects are compact with little resolved velocity information and highly affected by beam smearing which systematically increases velocity dispersion within the PSF. In our sample, which has median spatial resolution of $0\farcs4 \sim 3.3$ kpc at $z\sim1$ after accounting for magnification, we find an upper limit of 3/25 dispersion dominated galaxies. This is consistent with the values of $6-9\%$ from AO surveys \citep{Newman2013a,Jones2010a,Livermore2015} and lower than the fraction in \kmosd and KROSS ($\sim17\%$ at $z\sim1-2$): lensing is able to resolve kinematics on a comparable scale to AO. 

The three dispersion dominated galaxies in our sample are extremely different from each other - one is very low mass (MACS0416\_706, $\log \mstar = 8.31$), one is intermediate mass (MACS0416\_94, $\log \mstar = 9.88$) and the other high mass (MACS1149\_1644, $\log \mstar = 10.48$). The velocity dispersion of MACS1149\_1644 is the highest in the sample ($200\pm12$ km/s) and much more like dispersion dominated systems in previous work. This object has a low sSFR $\sim0.2\ssfrunit$ and could be a galaxy falling off the main sequence of star formation. MACS0416\_94 has a low measured velocity dispersion ($<38$ km/s after correcting for instrumental resolution and beam smearing) and relatively high sSFR $\sim2\ssfrunit$. MACS0416\_706 is a compact low mass galaxy at $z=1.35$ also with a very low measured velocity dispersion of $<36$ km/s, and high sSFR $\sim3\ssfrunit$ and may represent a new class of low mass compact dispersion dominated objects, forming an elliptical structure in situ, which are undetected in surveys lacking the increase in depth and resolution that gravitational lensing provides in KLASS.
\\
\section{Conclusions}\label{sec:conc}
We have presented the first results from KLASS, showcasing KMOS IFU spectroscopy of 32 gravitationally lensed galaxies at cosmic noon. Our key findings are: 

\begin{enumerate}

\item Emission line flux measured with KMOS is consistent with measurements of the same emission lines in the \HST G102 and G141 grisms. This is in contrast to recent follow-up of \HST grism-selected objects with slit-based spectrographs. Using simulated slits we find that slits recover only $\sim60\%$ of the flux compared to KMOS, and this fraction declines rapidly if the emission line is offset from the center of the slit.

\item In 25 of the 32 galaxies presented here we obtain high S/N kinematic maps, which show a diversity in kinematic structure. The majority of unresolved galaxies are in the field with the shortest integration time. The majority of our sample with resolved kinematics have $V_\textrm{max}/\sigma_0 > 1$ suggesting they are rotation dominated.

\item 3/25 of the resolved galaxies are classified as regularly ordered rotators with mean $V_\textrm{max}/\sigma_0 = 12.7\pm2.4$, similar to local disks, but existing only at stellar masses $>3\times10^{9} M_\odot$.

\item 13/25 of the resolved galaxies are classified as irregularly rotating systems. The mean $V_\textrm{max}/\sigma_0 = 5.5\pm0.5$ for these systems is lower than most disks in the local universe, indicating these are relatively turbulent `hot' disks. Trends in $V_\textrm{max}/\sigma_0$ are dominated by stellar mass.

\item With the power of lensing, we have resolved kinematics in galaxies with stellar masses below $>3\times10^{9} M_\odot$, none of which are regularly rotating and which have high sSFRs, indicating ongoing kinematic and morphological changes. 5 galaxies in our sample have stellar mass below $6.3\times10^8 M_\odot$, the lowest stellar mass objects at $z>1$ ever observed with resolved kinematics.

\item We find a lower fraction of dispersion dominated systems compared to comparable surveys in blank fields. This is likely because the enhanced spatial resolution from lensing allows us to resolve velocity gradients in more compact systems consistent with results from surveys using AO.

\end{enumerate}

Using the power of cluster lensing we have been able to efficiently resolve kinematics in objects at lower stellar masses than comparable multi-object IFU surveys, and at higher spatial resolution. We have seen a diversity in kinematic features for our sample, but find that only the highest mass objects form regular rotating disks at $z\simgt1$, whilst lower mass galaxies are irregularly rotating or likely to be involved in mergers.

When the full survey is complete KLASS will provide kinematics of $\sim60$ galaxies at cosmic noon. Benefiting from 10 hour final integration times we expect be able to resolve velocity gradients at the edges of the objects and produce rotation curves to large radii for our wide range of stellar mass.


\acknowledgments

This paper is based on observations made with ESO Telescopes at the La Silla Paranal Observatory under programme IDs 095.A-0258 and 196.A-0778. The authors thank the ESO KMOS team for carrying out our observations and answering our questions about the design and reduction of this Large Program. We thank Trevor Mendel for his code to correct the detector read-out bias and for insights into KMOS data reduction.

This work was supported by the \HST~GLASS grant GO-13459. C.M. acknowledges support by NASA Headquarters through the NASA Earth and Space Science Fellowship Program Grant 16-ASTRO16F-0002. T.T. acknowledges support by the Packard Foundation through a Packard Fellowship. T.J. acknowledges support provided by NASA through Program HST-HF2-51359 through a grant from the Space Telescope Science Institute, which is operated by the Association of Universities for Research in Astronomy, Inc., under NASA contract NAS 5-26555. T.M. acknowledges support from the Japan Society for the Promotion of Science (JSPS) through JSPS research fellowships for Young Scientists. B.V. acknowledges the support from an Australian Research Council Discovery Early Career Researcher Award (PD0028506).

This research made use of the following open-source packages for Python and we thank the developers of these:
Astropy \citep{Astropy},
SciPy libraries \citep[][including numpy and matplotlib]{Scipy}
and
PyFITS which is a product of the Space Telescope Science Institute, which is operated by AURA for NASA.

\bibliographystyle{apj}
\bibliography{library}

\begin{appendix}
\section{Kinematic Maps}\label{app:maps}

\begin{figure*}[t!] 
\centering
\figurenum{A1}
\includegraphics[width=0.95\textwidth]{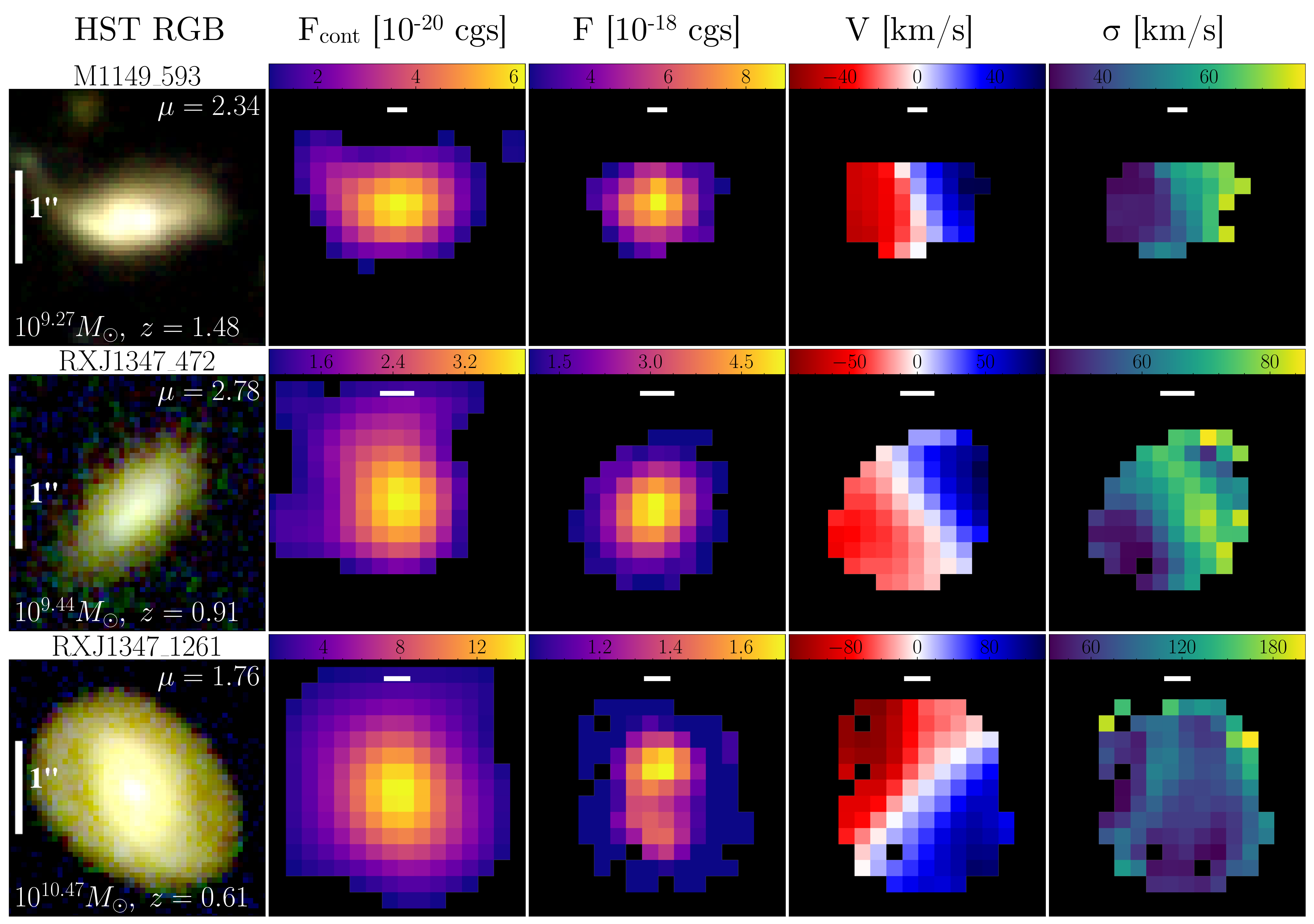} 	 
\caption{The \HST RGB composite images, KMOS YJ continuum flux, 2D emission line spectra and velocity maps for the \textit{regular rotators} class 1 of galaxies in KLASS, ordered by stellar mass. All maps are on the same spatial scale. The vertical white bar indicates $1\arcsec$ in the image plane and the horizontal white bar indicates 1 kpc in the source plane (or 100 pc if indicated in cases of high magnification).}
\label{fig:postage1}
\end{figure*}

\begin{figure*}[t!] 
\figurenum{A3}
\centering
\includegraphics[width=0.95\textwidth]{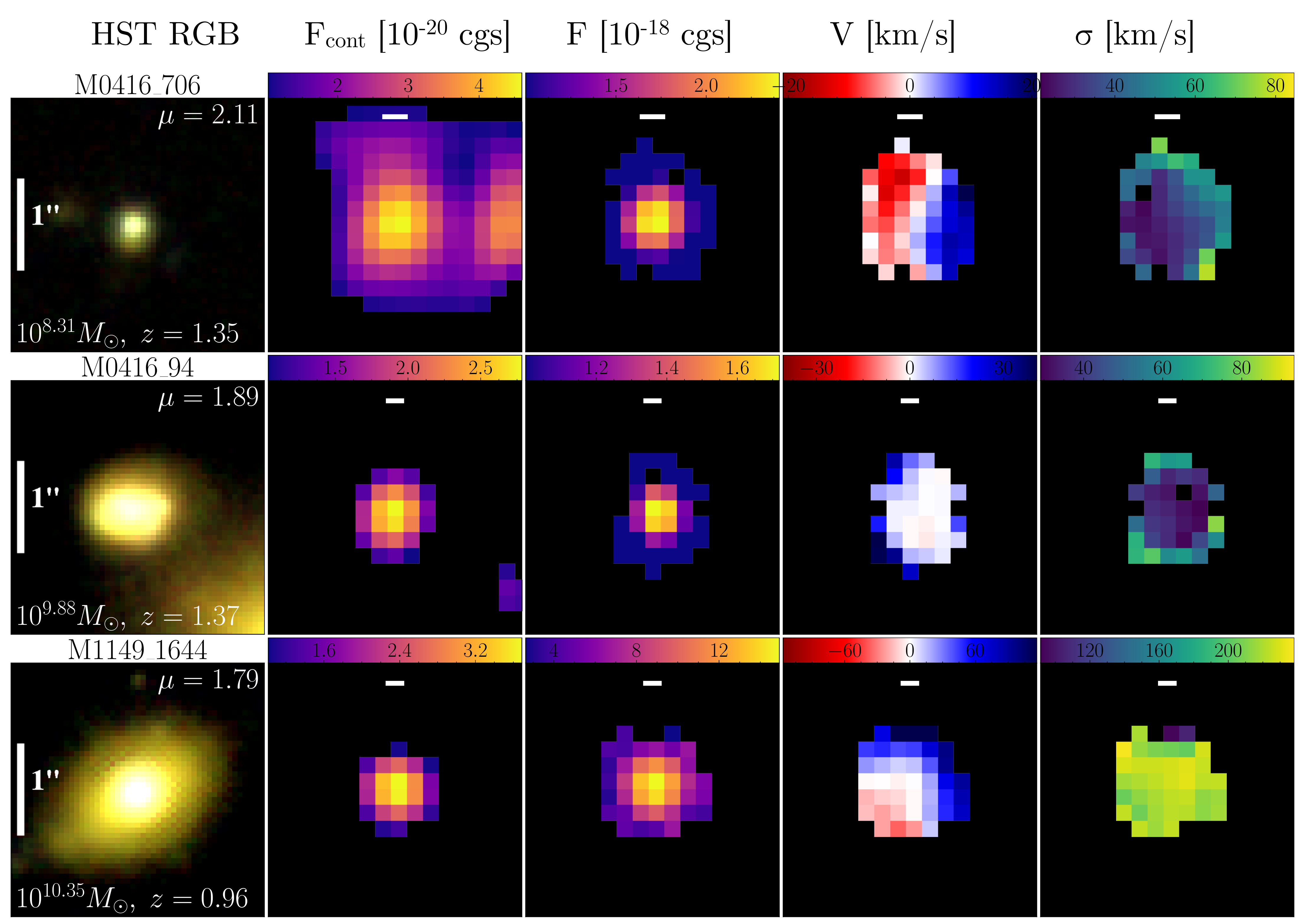} 	 
\caption{The \HST RGB composite images, KMOS YJ continuum flux, 2D emission line spectra and velocity maps for the \textit{dispersion dominated} class 3 of galaxies in KLASS, ordered by stellar mass. All scales and labels are the same as in the above figures.}
\label{fig:postage4}
\end{figure*}

\begin{figure*}[t!] 
\figurenum{A4}
\centering
\includegraphics[width=0.93\textwidth]{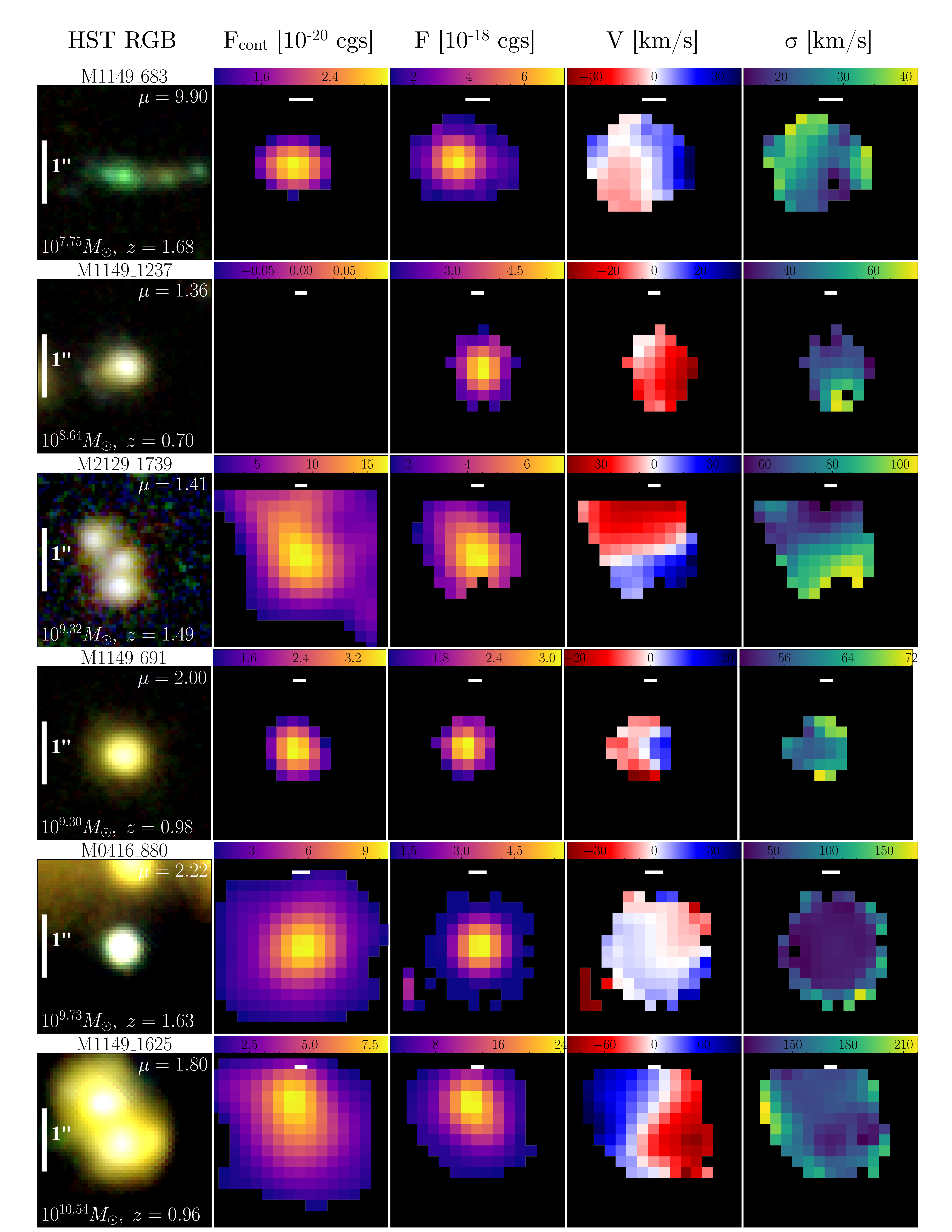} 	 
\caption{The \HST RGB composite images, KMOS YJ continuum flux, 2D emission line spectra and velocity maps for the \textit{mergers and unknown} class 4 of galaxies in KLASS, ordered by stellar mass. All scales and labels are the same as in the above figures.}
\label{fig:postage5}
\end{figure*}

The resolved high S/N kinematics maps for our sample are shown in Figures~\ref{fig:postage1}-\ref{fig:postage5}. The procedure for producing the maps is described in Section~\ref{sec:res_kinematics}.

We also show the \HST RGB images for each galaxy. For MACS0416 and MACS1149 RGB images are produced using the HFF data \citep{Lotz2016} in \HST Wide Field Camera 3 (WFC3) filters F606W, F125W and F160W. For MACS2129 and RXJ1347 we use F606W, F125W and F160W data from CLASH \citep{Postman2011}.

\begin{figure*}[t!] 
\centering
\figurenum{A2}
\includegraphics[width=0.8\textwidth]{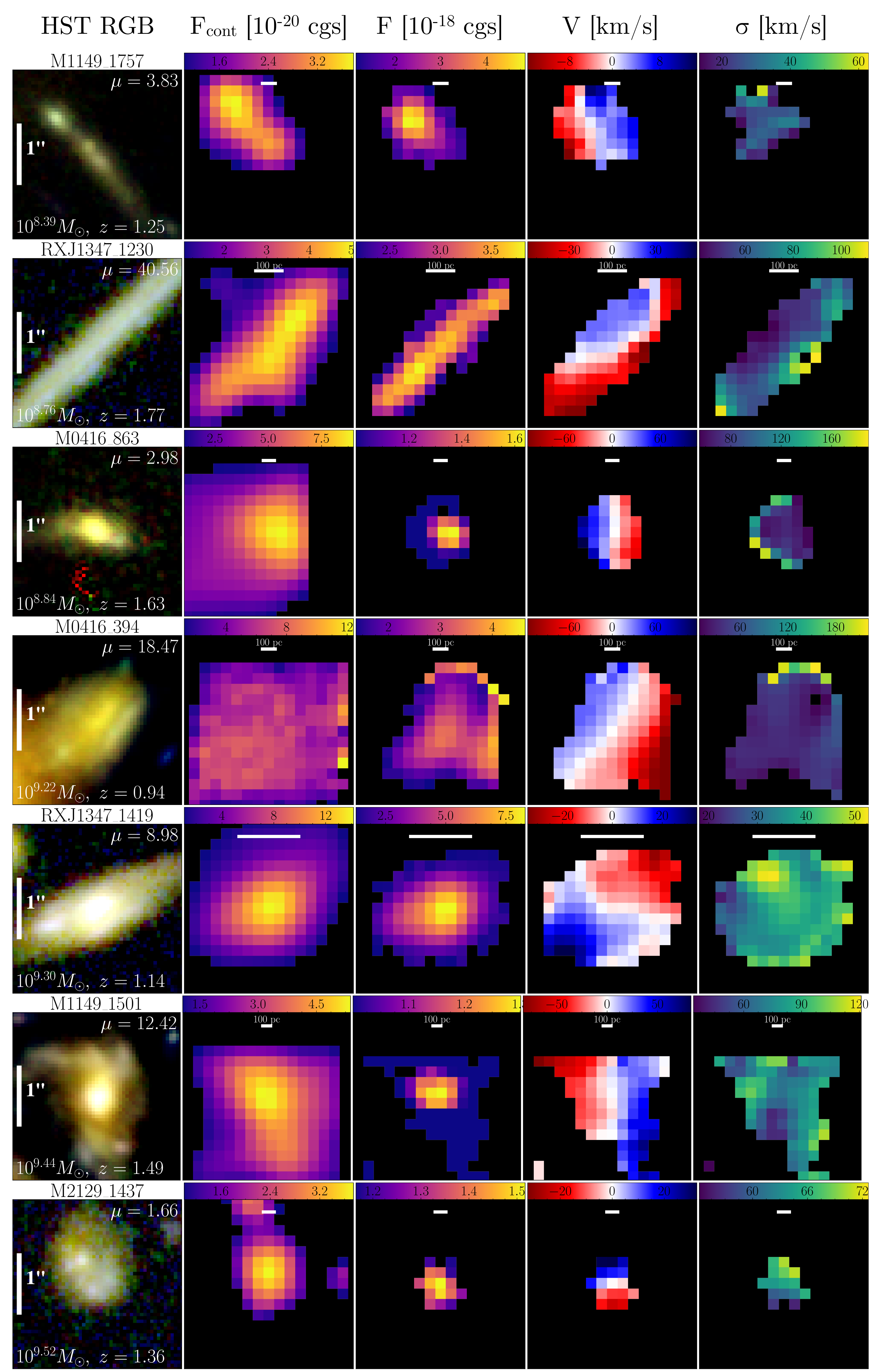} 	 
\caption{The \HST RGB composite images, KMOS YJ continuum flux, 2D emission line spectra and velocity maps for the \textit{irregular rotators} class 2 of galaxies in KLASS, ordered by stellar mass. All scales and labels are the same as in the above figures.}
\label{fig:postage2}
\end{figure*}

\begin{figure*}[t!] 
\figurenum{A2}
\centering
\includegraphics[width=0.95\textwidth]{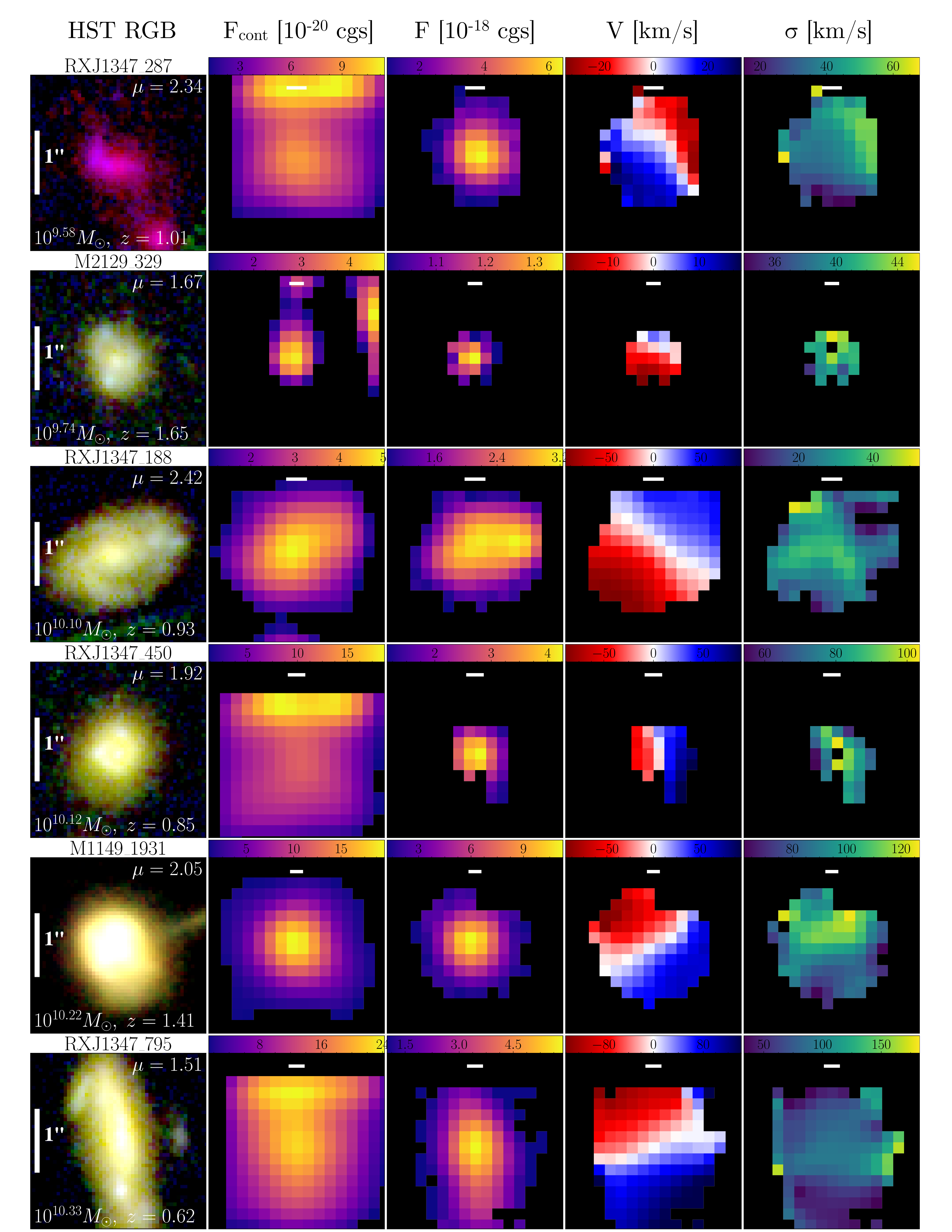} 	 
\caption{\textit{(cont.)}}
\label{fig:postage3}
\end{figure*}

\section{Rotation Curves}\label{app:rotcurves}

Rotation curves for rotating disks in our sample are presented in Figure~\ref{fig:rotcurves_reg}-\ref{fig:rotcurves_irreg1} below. Rotation curves are obtained using the procedure outlined in Section~\ref{sec:res_kinematics}.

\begin{figure*}[t!] 
\centering
\figurenum{B1}
\includegraphics[width=0.99\textwidth]{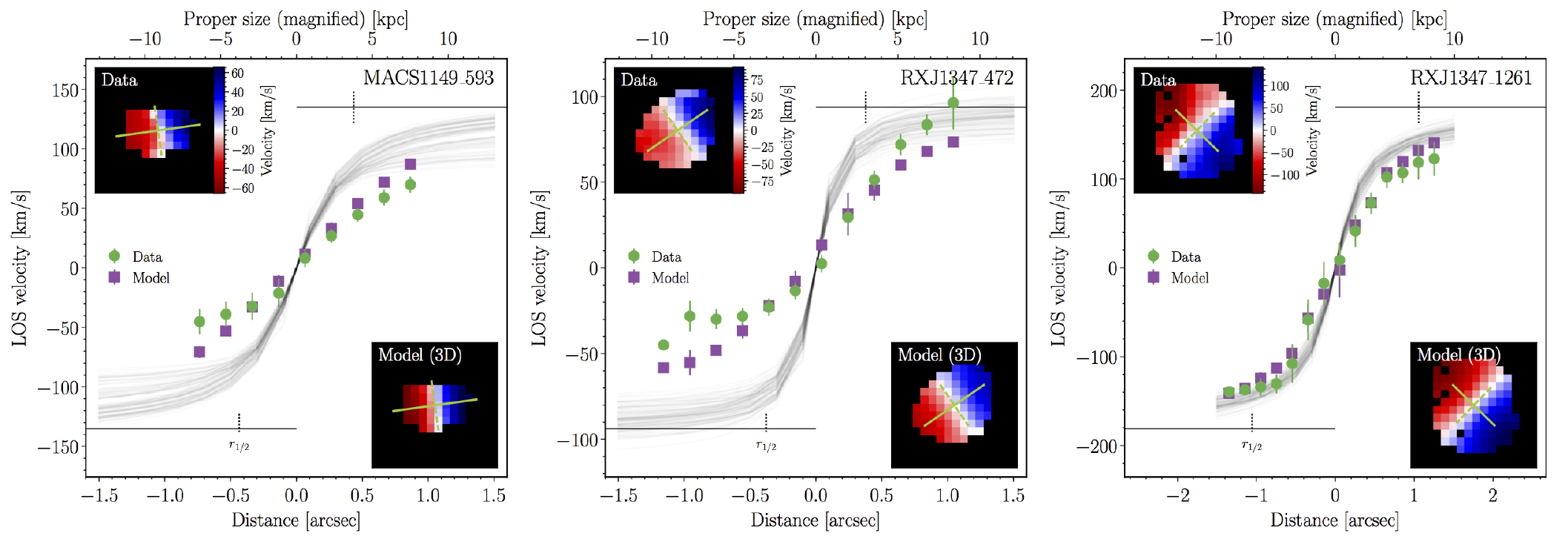} 	 
\caption{Rotation curves for the \textit{regular rotators} class 1 of galaxies in KLASS. We plot the 2D measured velocity maps and convolved model velocity maps from 3D fitting to the data cube via \galpak. We plot the rotation curve extracted from a one pixel slit along the kinematic major axis as indicated by the green solid lines on the maps. We plot points from the measured map (green circles) and model map (purple squares, convolved with the 3D PSF and LSF kernel - see Section~\ref{sec:res_kinematics}) and 100 sample line of sight model rotation curves drawn from the MCMC chain (gray lines). Solid horizontal black lines show the best-fit observed line-of-sight $V_\textrm{max}$. Dotted black vertical lines show the inferred best-fit half-light radii from \galpak or from GALFIT to the HST images in the 2D kinematic fits if they are not strongly lensed. Rotation curves are measured beyond $r_{1/2}$ for the majority of objects.}
\label{fig:rotcurves_reg}
\end{figure*}

\begin{figure*}[] 
\centering
\figurenum{B2}
\includegraphics[width=0.99\textwidth]{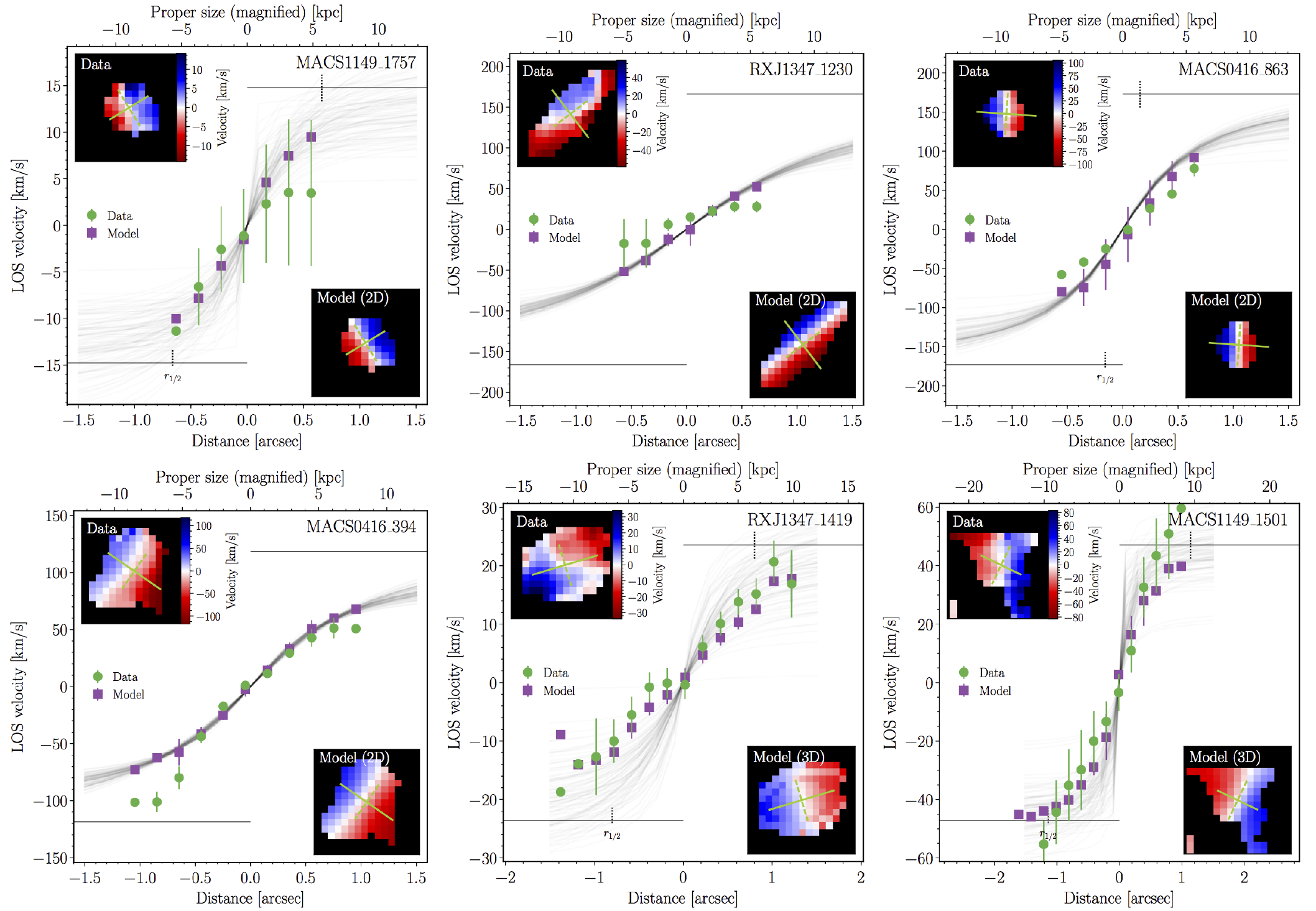} 	 
\caption{Rotation curves for the \textit{irregular rotators} class 2 of galaxies in KLASS, plotted in the same way as Figure~\ref{fig:rotcurves_reg}. Models are either 3D (from \galpak) or the 2D method described in Section~\ref{sec:res_kinematics}.}
\label{fig:rotcurves_irreg1}
\end{figure*}

\begin{figure*}[] 
\centering
\figurenum{B2}
\includegraphics[width=0.99\textwidth]{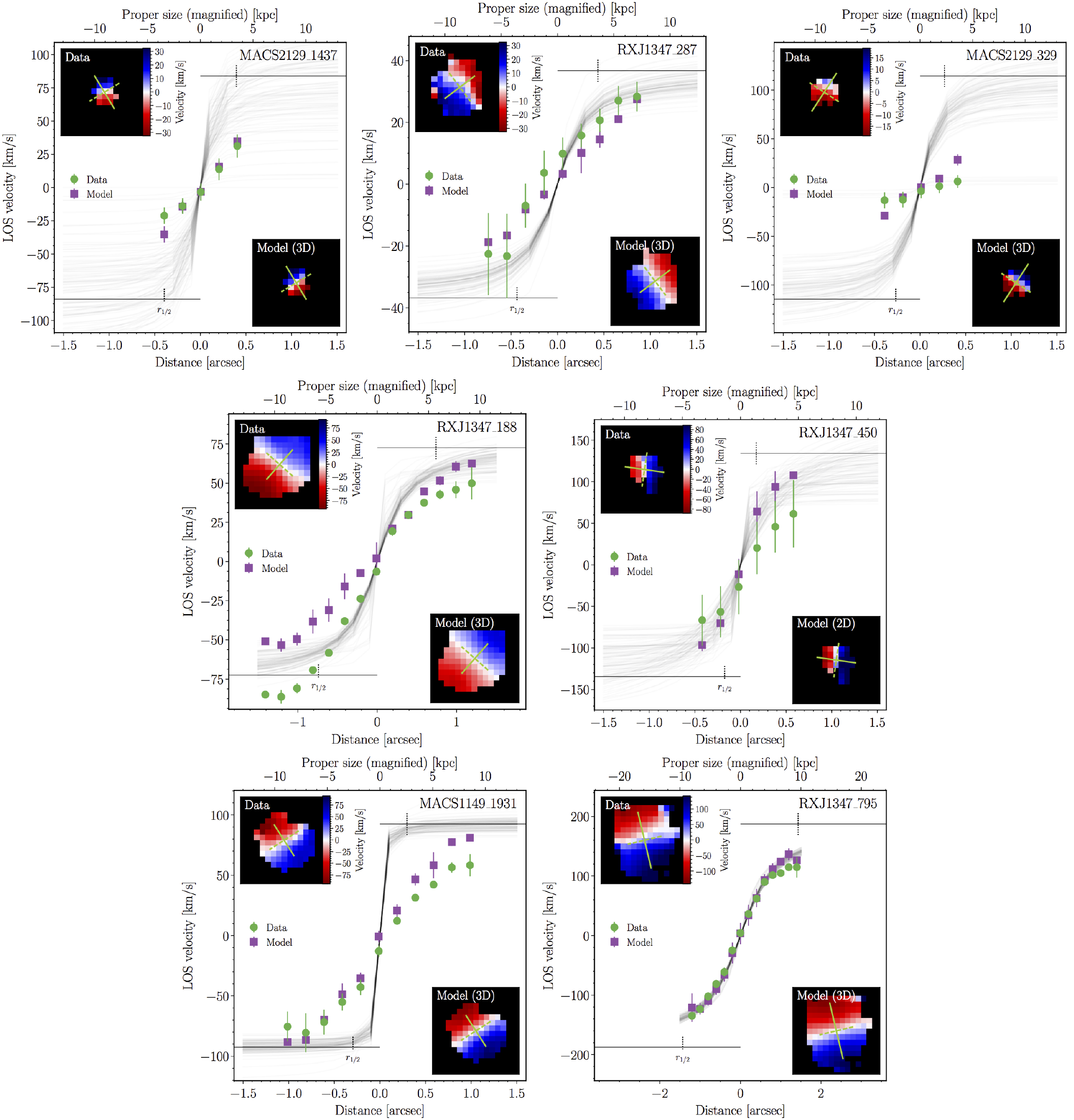} 	 
\caption{\textit{(cont.)}}
\label{fig:rotcurves_irreg2}
\end{figure*}

\end{appendix}

\end{document}